\documentclass[preprintnumbers,nofootinbib,noshowpacs,eqsecnum,prd]{revtex4}

\usepackage{graphicx,epsfig}
\usepackage{amsmath,amssymb,bbold,color}
\usepackage{epsfig,bbm}
\usepackage{url}
\usepackage{rotating}
\usepackage{hhline}

\graphicspath{ {Plots/} }
\makeatletter
\renewcommand{\thesection}{\arabic{section}}
\renewcommand{\p@subsection}{}
\renewcommand{\thesubsection}{\arabic{section}.\arabic{subsection}}
\makeatother

\begin{document}


\thispagestyle{empty}

\begin{center}

{\large\sc {\bf Multiple Higgs-Portal and Gauge-Kinetic Mixings}}
\\[3.5em]
{\normalsize
{\sc
S.~Y.~Choi$^{1}$, C.~Englert$^{2}$, and P.~M.~Zerwas$^{3}$
}
}

\vspace*{0.75cm}

{\it
$^1$ Department of Physics, Chonbuk National University, Jeonju 561-756,
     Republic of Korea

$^2$ SUPA, School of Physics and Astronomy, University of
  Glasgow, Glasgow, G12 8QQ, United Kingdom

$^3$ Theory Group, Deutsches Elektronen-Synchrotron DESY, D-22603
Hamburg, Germany
}

\vskip 0.1cm
(\today)
\end{center}

\vspace*{1.5cm}

\begin{abstract} {\noindent We develop a phenomenological formalism
    for mixing effects between the Standard Model and hidden-sector
    fields, motivated by dark matter in the Universe as well as string
    theories. The scheme includes multiple Higgs-portal interactions
    in the scalar sector as well as multiple gauge-kinetic mixings in
    the abelian gauge sector. While some of the mixing effects can be
    cast in closed form, other elements can be controlled analytically
    only by means of perturbative expansions in the ratio of standard
    scales over large hidden scales. Higgs and vector-boson masses and
    mixings are illustrated numerically for characteristic processes.
  }
\end{abstract}

\maketitle

\setcounter{page}{1}
\setcounter{footnote}{0}

\vspace{8mm}


\section{Basics}

\noindent
A large fraction of matter in the Universe is invisible
\cite{Ade:2013zuv}.  This hidden sector may have structures at least
as complex as matter and interactions in the visible Standard Model
[SM] sector (see e.g. \cite{McCullough:2013jma}), unlike one-component
theories as realized in supersymmetry approaches to the dark
sector. Opportunities to explore structures in this hidden sector are
offered by mixing effects with fields of the SM.  Couplings between
the two sectors are provided by Higgs-portal interactions
\cite{Hiportal, Hportal_kinmix,Hiportal2,hiportalnlo,Andersen:2013rda}
and gauge-kinetic mixings
\cite{Hportal_kinmix,Holdom,kinmix}. Higgs-portal interactions couple
the invariant bilinear product of the Higgs field in the Standard
Model with Higgs SM-singlets in the hidden sector [HS], the strength
of the interaction measured by $\eta$. Kinetic mixing couples abelian
hypercharge $B,V$-field tensors in both sectors with strength $s$.
The coupling $\eta$ can be varied in the analysis in a large range
\cite{NN22}, $|\eta| \leq 1$, and the kinetic coupling $s$ in the most
general scenario in a similar range $|s| \leq 1$, depending strongly
however on the underlying microscopic picture of the mixing
mechanism.{\footnote{We are very grateful to J.~Jaeckel for valuable
    advice on field- and string-theoretic models of kinetic mixing and
    consequences for the potential range of $|s|$.}}  If in
field-theoretic models loops of mediators link the hypercharge $B,V$
fields in the two sectors, the size of $|s|$ is expected to be
restricted to $|s| \sim g_{_{\rm SM}} g_{_{\rm DM}} / 6 \pi^2 \times
{\rm mass}\;{\rm logs} \lesssim 10^{-2}$, set in detail by the gauge
couplings, the hypercharges and the masses \cite{Holdom}.  On the
other hand, the kinetic coupling as mediated by string states is less
stringently restricted in general, and it can be quite large depending
on the type of string theory realized, see e.g. \cite{Jaeckelpaper}.
We will not delve into building a detailed model of the joined
[SM]$\oplus$[HS] system, nor of dark-matter candidates.  However, the
HS system may be taken as a SM-type system \cite{Barbieri}, properly
extended to accommodate the dark-matter properties. The lightest
fermion field, among others, could be a stable candidate in such a
scenario for the dark-matter field.  The hidden-sector scales will be
assumed of TeV size, contrasting approaches in which fields are
assumed super-light. The present analysis aims at exploring the
potential of high energy colliders, LHC
and LC, for shedding light on the structure of a heavy hidden sector. \\

The SM electroweak gauge interaction of the $B$-field is unaltered by
kinetic mixing to leading order and the procedure of high-precision SM
analyses is modified only to second order. Also, the Higgs-portal
interactions have been introduced in renormalizable form. As a result,
well-explored low-energy physics is not dramatically affected, and
changes are just
restricted to potentially small corrections. \\

It turns out that the re-diagonalization of the gauge sector after
switching on kinetic mixing is surprisingly straightforward and the
results can largely be presented in transparent analytic
form. However, the re-diagonalization of the effective mass matrix of
the gauge fields, parallel to the re-diagonalization of the mass
matrix in the Higgs sector, can in general be carried out analytically
only by expanding mass eigenvalues and field eigenvectors in the ratio
of SM masses over dark masses, the latter assumed to be large, i.e.
in the TeV regime \cite{analytic_expansion}. \\

First we will discuss both Higgs-portal mixing and gauge-kinetic
mixing quite generally.  Thereafter we will turn to numerical examples
for Higgs and vector fields in a hidden sector coupled either by one or
two links to the SM, i.e. $1\oplus 1$ and $1\oplus 2$ mixing
scenarios. Specifically we will address the problem of how to encircle
parameters such that the structure of the combined system can be
tested.  We will analyze the problem of how to extract, at LHC and LC,
the basic mass parameters
and the mixings between the hidden sector and the measurable SM. \\


\subsection{Higgs Portal : Multiple Couplings}

\noindent
In the extended Higgs sector we will assume that the SM iso-doublet
Higgs field $\phi$ is complemented by a set of $n$ complex scalar
fields $S_i$ that generate the hidden gauge boson masses, including the
$n$ U(1) gauge bosons $V= \{V_1,\cdots,V_n \}$, which are mixed with
the hypercharge SM gauge boson. Expanding the fields about the minimum
of the scalar potential, the real SM Higgs field $H_0$ and $n$
hidden-sector Higgs fields $H= \{H_1,\cdots,H_n \}$ will emerge,
corresponding to scalars located primarily in the visible SM sector
and the invisible hidden sector, respectively. The two sectors are
coupled weakly and they are linked
by bi-linear quartic couplings, leaving the system renormalizable. \\

Cast into the standard formalism of spontaneous symmetry breaking, the
Higgs potential is introduced as
\begin{equation}
{\mathcal{V}}_{\cal H} =  \left[\, \mu^2_0 |\phi|^2 + \lambda_0 |\phi|^4 \, \right]
                 + \sum_{i=1}^n\,\left[\,\mu^2_i |S_i|^2 + \lambda_i |S_i|^4\,\right]
                 + \sum_{i=1}^n\, \eta_i |\phi|^2 |S_i|^2 \,.
\end{equation}
The $\mu^2$ parameters are negative, shifting the ground states to
non-zero vacuum values, and the $\lambda$ parameters are positive to
stabilize the system. As we focus primarily on the direct coupling of
the SM Higgs field and the hidden-sector Higgs fields, quartic
interactions among the hidden-sector Higgs fields can be ignored to
leading order in this context if the couplings of the mixing terms are
taken to be small. Their impact on any physical observables will be
suppressed when passing the Higgs-portal. [For illustration, the
analysis including small quartic mixing
in the hidden sector is summarized in Appendix~\ref{append:exhiggs}.\\

Introducing the vacuum Higgs values $v$ at the
minimum of the potential,
\begin{equation}
  \phi_0 = \left(v_0 + H_{0c}\right) / \sqrt{2}  \quad {\rm and} \quad
  S_i  = \left(v_i + H_{ic}\right) / \sqrt{2}
\end{equation}
where $\phi_0$ is the neutral component of the SM iso-doublet Higgs
field $\phi$ etc, they can be expressed by the potential parameters
after solving the minimum conditions:
\begin{eqnarray}
  v^2_0 \! &=& \! [-\mu^2_0 - \tfrac{1}{2} \sum_{k=1}^n \eta_k v^2_k] / \lambda_0
              \ \ \rightarrow \ \
              \frac{-\mu^2_0/\lambda_0 + \tfrac{1}{2} \sum \eta_k \mu_k^2/\lambda_k \lambda_0}
                   { 1 - \tfrac{1}{4} \sum \eta_k^2 / \lambda_k \lambda_0 } \\[1mm]
  v^2_i \! &=& \! [-\mu^2_i - \tfrac{1}{2} \eta_i v^2_0]/ \lambda_i
              \ \ \qquad \rightarrow \ \
              \frac{-\mu_i^2/\lambda_i + \tfrac{1}{2} \eta_i \mu_0^2/\lambda_0 \lambda_i
                    -\tfrac{1}{4} \sum [\eta_i \mu_k^2 - \eta_k \mu_i^2 ] \eta_k / \lambda_k \lambda_0 \lambda_i }
              { 1 - \tfrac{1}{4} \sum \eta_k^2 / \lambda_k \lambda_0 }
                   \ \ \mbox{for} \ \  i = 1,... n    \,.
\end{eqnarray}
Inserting mutually $v_0$ into $v_i$ and v.v., the set of two equations
has been solved for $v_0$ and $v_i$ supplementing the original values
of the individual sectors, $v_0^2\leftarrow -\mu^2_0/\lambda_0$ and
$v_{i}^2 \leftarrow -\mu_i^2/\lambda_i$, before they are coupled by
$\eta_i$. \\

The bilinear kinetic terms and the mass terms of the physical Higgs
fields are described by the effective Lagrangian:
\begin{equation}
  \mathcal{L}_H
= \frac{1}{2} \; \partial
  \left( \begin{array}{c}
         H_0 \\
         H
         \end{array} \right)^{\!\!T}_{\!c}
        \!\!
        \,\partial
        \left( \begin{array}{c}
                  H_0 \\
                  H
                  \end{array} \right)_{\!c}
- \frac{1}{2}\,
  \left( \begin{array}{c}
         H_0 \\
         H
         \end{array}  \right)^{\!\!T}_{\!c}
         \!\!
  {\mathcal{M}}_{Hc}^2
  \left( \begin{array}{c}
         H_0 \\
         H
         \end{array}            \right)_{\!c}
\quad\; {\rm with} \quad\;
  {\mathcal{M}}_{Hc}^2
= \left(\begin{array}{cc}
        M^2_{0c}   & X^T  \\
        X          & M_c^2
        \end{array}       \right)           \,.
\end{equation}
The parameter $ M^2_{0c}$ denotes the mass of the SM Higgs boson in
the current basis, the matrix $M^2_c$ denotes the $n \times n$ mass
matrix of the Higgs bosons in the hidden sector [for simplicity assumed
to be diagonal], and the $n$-dimensional column vector $X$ accounts
for the couplings of the scalars in the SM and the hidden sectors:
\begin{eqnarray}
   \label{eq:m1}
    M^2_{0c} &=& 2 \lambda_0 v^2_0            \\
   \label{eq:m2}
    M^2_c    &=& {\rm diag}\left( 2\lambda_1 v^2_1,\cdots,
                                  2\lambda_n v^2_n\right)    \\
   \label{eq:m3}
    X^T      &=& (\eta_1 v_0 v_1,\cdots,\eta_n v_0 v_n)
\end{eqnarray}
expressed by the basic parameters of the potential. [In the general
case with quartic terms of hidden-sector Higgs fields, the matrix
$M^2_c$ is non-diagonal but symmetric so that it can be diagonalized
by an orthogonal transformation. The mixing vector $X$ is changed
slightly as a result, see Appendix~\ref{append:exhiggs}.]\\

The mass matrix in the current $[c]$ representation will be
transformed to the diagonal mass matrix in the mass $[m]$
representation by applying the orthogonal transformation
$\mathcal{O}_H$:
\begin{eqnarray}
 \mathcal{M}^2_{Hc}\, \Rightarrow\, {\mathcal{M}}^2_{Hm}
  \, =\, \mathcal{O}_H \mathcal{M}^2_{Hc} \mathcal{O}^T_H
  \, =\, {\rm diag}\,({M}_{0m}^2,{M}_m^2)
\end{eqnarray}
while the Higgs fields transform as
\begin{equation}
  \left( \begin{array}{c}
         H_0 \\
         H
         \end{array}    \right)_{\!c}
=\; \mathcal{O}^T_H
  \left( \begin{array}{c}
         H_0  \\
         H
         \end{array}    \right)_{\!m} \,.
\label{eq:O_H}
\end{equation}
\\

For $n = 2$ and higher, the eigenvalues and eigenfunctions of the mass
matrix cannot be written in closed or transparent form anymore. But
they can be expanded consistently for small mixing up to second order
in the expansion parameter $X$~\cite{analytic_expansion}:
\begin{eqnarray}
 {M}_{0m}^2 &=& M^2_{0c} - X^T (M_c^2 - M^2_{0c})^{-1} X         \\[2mm]
 {M}_m^2    &=& M^2_c + \tfrac{1}{2} {\rm diag}
                        \left\{ X X^T ,\, (M_c^2 - M^2_{0c})^{-1} \right\}
\label{Meigen}
\end{eqnarray}
while the orthogonal transformation matrix up to second order reads
\begin{eqnarray}
  \mathcal{O}_H
= \left( \begin{array}{cc}
         1-\frac{1}{2} {\Omega}^T_H \Omega_H &
          {\Omega}^T_H   \\
         -\Omega_H                           &
          1-\frac{1}{2} \Omega_H {\Omega}^T_H
         \end{array}  \right)
\quad\; {\rm with} \quad\;
\Omega_H =-( M^2_c - M^2_{0c} )^{-1} X \,.
\end{eqnarray}
The corrections of the heavy hidden Higgs masses $M^2_m$ are in general
not diagonal. However, re-diagonalization gives rise to changes of the
eigenvalues and mixing matrices only beyond the order considered so
that the off-diagonal elements can simply be truncated [see the proof
in the Appendix]. \\

By assuming the mass parameters in the hidden sector to be heavy
compared with the SM sector and the mixing parameters, the norm of the
vector $|| \Omega_H || \sim ||X||/||M^2||$ is small and serves as an
expansion parameter, in parallel with $||M_0^2||/||M^2||$.  The SM
Higgs mass parameter $M^2_{0c} \Rightarrow {M}^2_{0m}$ however could
be modified sizably if the mixing parameter is not much smaller than
the SM parameter [consistent with the expansion].  To lowest order,
the modifications of the masses and the mixing matrix,
\begin{eqnarray}
 {M}^2_{0m} &\simeq& M^2_{0c} - X^T (M_c^2)^{-1} X     \\[2mm]
 {M}_m^2   &\simeq& M_c^2
\end{eqnarray}mcc
and
\begin{equation}
  \mathcal{O}_H
\simeq
  \left( \begin{array}{cc}
         1          & {\Omega}^T_H               \\[1mm]
         - \Omega_H & 1
         \end{array}  \right)
\quad {\rm and} \quad
  \left\{ \begin{array}{l}
          \, H_{0m}   \simeq \quad\!\!  H_{0c} \;\;\quad\;+\;
          \Omega^T_H H_c  \\[2mm]
          \, H_{m}  \;\,\simeq   - H_{0c} \Omega_H \, +\; H_c
          \end{array}\right.
\end{equation}
are particularly simple. This set of transformations of the
wave-functions generates the reduced couplings of the SM-type Higgs boson
and the couplings of the hidden Higgs bosons
with the SM gauge and matter fields to first order in  the mixing. \\

In the $1 \oplus 1$ Higgs scenario the solution can be reconstructed
analytically without any expansion \cite{NN22,Bertolini:2012gu}. The
mixing matrix is the standard 2$\times$2 matrix built up by $\sin\chi$
and $\cos\chi$ of the rotation angle $\chi$. After reducing the vector
$\Omega_H$ to the small mixing angle $\Omega_H \simeq \chi$, the above
relations are readily recovered from the $(1+n)\times (1+n)$ system, as
will be recalled later. \\

Starting from the lowest order, masses and mixings can iteratively be
constructed to arbitrary order in the expansion parameter
$\epsilon \sim ||M^2_{0}|| / ||M^2||,\, ||X ||/ ||M^2||$, both of
which are small for large masses in the hidden sector, compared to
SM masses and mixings. The perturbative recursion formulae are derived
in the Appendix. \\


\subsection{Kinetic Mixing}

\noindent
The interaction between the SM hypercharge $B$-field and the set of $n$
gauge $\{ V_1,\cdots,V_n \}$ fields concentrated in the hidden sector,
is described by the Lagrangian
\begin{eqnarray}
\mathcal{L}_V
   \!=\! -\frac{1}{4}\left( \begin{array}{c}
                      \tilde{W} \\
                      \tilde{B} \\
                      \tilde{V}
                      \end{array}         \right)^{\!\!T}_{\! c}\!
                      [1+\mathcal{S}\,] \!
               \left( \begin{array}{c}
                      \tilde{W} \\
                      \tilde{B} \\
                      \tilde{V}
                      \end{array} \right)_{\! c}
   \! + \,\frac{1}{2} \left( \begin{array}{c}
                            {W} \\
                            {B} \\
                            {V}
                            \end{array}         \right)^{\!\!T}_{\! c}
                            \! {\mathcal{M}}_{Vc}^2
                     \left( \begin{array}{c}
                            {W} \\
                            {B} \\
                            {V}
                            \end{array} \right)_{\! c}
\quad \mbox{with}\quad
\mathcal{S}\! =\! \left( \begin{array}{lll}
                     0  & 0  & 0             \\
                     0  & 0  &  s^T          \\
                     0  & s  &  0
                     \end{array}                \right)
\end{eqnarray}
where ${\mathcal{M}}^2_{Vc}$ is the current $(2+n)\times (2+n)$
gauge-boson mass matrix given by
\begin{eqnarray}
  {\mathcal{M}}^2_{Vc} \!  =\! M^2_{Z_c} \left( \begin{array}{ccc}
      c^2_W    & -c_W s_W    &  0  \\
      -c_W s_W  &  s^2_W      &  0  \\
      0        &   0         & \Delta
                             \end{array} \right) \quad {\rm with} \quad
                                                 \Delta = M^2_{V_c}/M^2_{Z_c}
\end{eqnarray}
in terms of the mass parameter $M_{Z_c} = \sqrt{g^2+g'^2}\, v_0/2$,
the sine/cosine of the electroweak mixing angle $s_W=\sin\theta_W$
etc, as well as the $n\times n$ dimensionless matrix $\Delta$,
after including the mixing of the SM neutral iso-spin $W$-field and
the hypercharge $B$-field due to electroweak symmetry breaking.  While
the field vectors are denoted by $V=\{ V_1,\cdots,V_n \}$, the field
tensors are denoted by $\tilde{V}= \{\tilde{V}_1,\cdots,\tilde{V}_n
\}$; $s$ is the $n$-dimensional vector accounting for the kinetic
$B$-$V$ mixings. For the sake of notational simplicity we have
refrained from introducing mixing among the fields in the hidden
sector, which can easily be added . \\

Applying an ${\rm SL}(2+n,{\rm R})$ matrix transformation, consisting
of a kinetic transformation [KT] $\mathcal{Z}$ and a rotation, to the gauge
fields, the kinetic mixing of the field strengths can be absorbed in
the redefinition of the fields. Thereafter, the mass matrix must be
diagonalized by a matrix split into the block-diagonalization matrix
${\mathcal{O}}_V$ and the rotation matrix $\mathcal{U}_d$
re-diagonalizing the mass submatrix in the hidden sector. The
block-diagonalization can be performed only approximately; the
expansion parameters being $||s|||$ and $||M^2_Z / M^2_V||$, with the
norms assumed to be small. The final result of this procedure can be
written as follows:
\begin{equation}
    \mathcal{L}_V
= 
    - \frac{1}{4} \left( \begin{array}{c}
                       \tilde{A} \\
                       \tilde{Z} \\
                       \tilde{V}
                       \end{array}         \right)^T_{\! m}
                        {1}_N                                 \,
                 \left( \begin{array}{c}
                        \tilde{A} \\
                        \tilde{Z} \\
                        \tilde{V}
                        \end{array} \right)_{\! m}
   + \,\frac{1}{2}\, \left( \begin{array}{c}
                            {A} \\
                            {Z} \\
                            {V}
                            \end{array}     \right)^T_{\! m}
                        \! {\mathcal{M}}^2_{Vm}            \,
                     \left( \begin{array}{c}
                            {A} \\
                            {Z} \\
                            {V}
                        \end{array} \right)_{\! m} \; \ \ \mbox{with}\ \  \;\;
\left( \begin{array}{c}
        W \\
        B \\
        V
        \end{array}         \right)_{\! c}
 = \; \mathcal{Z}^T\, \mathcal{O}^T_V \, \mathcal{U}^T_d
   \left( \begin{array}{c}
          A \\
          Z \\
          V
          \end{array}       \right)_{\! m}
\label{eq:BB}
\end{equation}
leading to the massless photon field $A_m$, and the massive vector
fields $Z_m$ and $V_m$. The KT matrix and the
block- and re-diagonalization matrices read
\begin{eqnarray}
   \mathcal{Z}
 = \left( \begin{array}{ccc}
          s_W   &  c_W        &     0  \\
          c_W   & -s_W        &     0  \\
          0     & -\sigma\, s &  \sigma
          \end{array}     \right)
\,, \;\;
{\mathcal{O}}_V
\simeq \left( \begin{array}{ccc}
              1   &  0          &  0            \\
              0   &  1          & \Omega^T_V   \\
              0   & -\Omega_V   & {1}
              \end{array}  \right)
\ \  \mbox{and} \ \
\mathcal{U}_d = \left(\begin{array}{ccc}
                      1   &   0   &   0  \\
                      0   &   1   &   0  \\
                      0   &   0   &   U_d
                      \end{array}\right)
\label{eq:general_transformation}
\end{eqnarray}
with the $n$-dimensional column vector $\Omega_V$
\begin{eqnarray}
\Omega_V \,\simeq\, - s_W \left(M^2_{Z_c}/M^2_{V_c}\right) s
\end{eqnarray}
to lowest order in the mixing. [$\,\mathcal{U}_d$ actually proves
ineffective
to lowest order, as derived in the Appendix.] \\

The submatrix $\sigma$ in $\mathcal{Z}$ is a symmetric $n \times n$
matrix defined by the dyadic product $ss^T$ of the mixing parameters $s$,
\begin{equation}
\sigma = ({1} - ss^T\, )^{-\frac{1}{2}}
       = u^T\, \sigma_d\, u
\label{eq:stretch}
\end{equation}
which can be diagonalized by the $n\times n$ orthogonal matrix $u$,
generating the diagonal matrix $\sigma_d$.  The set of eigenvalues of
the dyadic matrix $ss^T$ consists of one non-zero value $||s||^2$
followed by $n-1$ zero values, giving rise to the $n\times n$ diagonal
matrix $\sigma_d = {\rm diag}\,
[(1-||s||^2)^{-\frac{1}{2}},1,\cdots,1]$.
The spectrum and the eigenvectors are derived in the Appendix. \\

Switching from the current $[c]$ basis to the mass $[m]$ basis, the mass
matrix is transformed, up to second order, to the diagonal
$(2+n)\times (2+n)$ mass matrix $\mathcal{M}^2_{Vm}$ in
Eq.$\,$(\ref{eq:BB}), representing the zero photon mass and the
non-zero mass eigenvalues up to second order as
\begin{eqnarray}
M^2_\gamma &=     & 0 \\
M^2_{Z_m}  &\simeq& M^2_{Z_c} - s^2_W s^T (M^4_{B_c}/M^2_{V_c})\, s
                    \;\; \to\;  M^2_{Z_c} \\
M^2_{V_m}  &\simeq& U_d M^2_{V_c} U^T_d +  \frac{1}{2}\, {\rm diag}
                       \left\{ ss^T,\, M^2_{V_c}\right\}
\end{eqnarray}
where $U_d$ diagonalizes the matrix $M^2_{V_c}$ up to the second-order
approximation of the mixing.\\

Special attention should be payed to the peripheral null-vector in the matrix
$\mathcal{Z}$. This form is essential to
keep the SM gauge interactions intact to leading order. This is apparent
by noting the covariant derivative which transform as
\begin{eqnarray}
 i\, D &=& i\,\partial - g T_3 W_c - g' Y B_c - g_V Y^T_V V_c
    \nonumber\\[2mm]
    &\simeq & i\,\partial
              - e Q A_m - g_Z (T_3-Y) Z_m
              - \left( g_V Y^T_V - g' Y s^T\right) V_m
\end{eqnarray}
with $Y_V= \{Y_{V1},\cdots,Y_{Vn} \}$ and up to linear approximation
in the kinetic mixing; as usual, $e= g s_W$, $Q=T_3+Y$ and $g_Z=
[g^2+g'^2]^{1/2}$. The coefficients of the $A$ and $Z$ fields are not
altered preserving the standard structures in the original
electroweak sector after electroweak symmetry breaking at this level.  \\

\section{$1 \oplus 1$ Analysis}
\label{sec:1+1an}

\noindent
The simplest example of portal models combines the SM with just one
new degree of freedom in the hidden sector. With some elements
worked out already a while ago, cf. Ref.$\,$\cite{Hiportal}, we extend
the analysis in this section at the level of phenomenology as well as
analytical solutions based on perturbative expansions. Note that the
Higgs and the gauge sectors are entangled, the connecting link being
the transformed vacuum expectation values of the Higgs fields affecting
the current mass parameters of the gauge fields. \\

\subsection{Higgs system}

\noindent
Specifying the notation in the previous sections, the physical Higgs
masses $M_{0m/1m}$ and the Higgs mixing angle $\chi$, derived from the
current Higgs mass matrix,
\begin{eqnarray}
  {\cal M}^2_H
= \left(\begin{array}{cc}
        2\lambda_0 v^2_0   & \eta_1 v_0 v_1 \\[1mm]
        \eta_1 v_0 v_1   & 2 \lambda_1 v^2_1
        \end{array}\right)
\end{eqnarray}
are given by
\begin{eqnarray}
&& M^2_{0m} = \lambda_0 v^2_0 + \lambda_1 v^2_1
            -\sqrt{(\lambda_1 v^2_1 -\lambda_0 v^2_0)^2
                    + (\eta_1 v_0 v_1)^2}
        \,\simeq\, 2\lambda_0 v^2_0
                 - \eta_1^2 v_0^2/2\lambda_1 \\
&& M^2_{1m} = \lambda_0 v^2_0 + \lambda_1 v^2_1
            +\sqrt{(\lambda_1 v^2_1 -\lambda_0 v^2_0)^2
                    + (\eta_1 v_0 v_1)^2}
            \,\simeq\, 2\lambda_1 v^2_1
                 + \eta_1^2 v_0^2/2\lambda_1 \\
&& \tan 2\chi =-\eta_1 v_0 v_1 /(\lambda_1 v^2_1-\lambda_0 v^2_0)
               \simeq -\eta_1 v_0 / \lambda_1 v_1
\end{eqnarray}
up to the second order approximation in $\eta$ for the masses.  It
should be noted that the leading mass corrections are not suppressed
by the large hidden scale $v_1$, in contrast to the mixing angle.  The
current Higgs fields $\{ H_{0c}, H_{1c} \}$ and the mass Higgs fields
$\{H_{0m},H_{1m} \}$ are related by the orthogonal transformation
${\cal{O}}_H^T$ :
\begin{eqnarray}
  \left( \begin{array}{c}
      H_0 \\
      H_1     \end{array} \right)_c
  = \left( \begin{array}{rr}
      \cos\chi & -\sin\chi  \\
      \sin\chi & \cos\chi  \end{array} \right) \,
  \left( \begin{array}{c}
      H_0 \\
      H_1     \end{array} \right)_m
\end{eqnarray}
where $\cos\chi$ can be assumed non-negative without loss of
generality.  These three observables can be exploited, in return, to
extract the individual vacuum parameters $\lambda_0 v^2_0,\lambda_1
v_1^2$ and $\eta_1 v_0 v_1$ according to
\begin{eqnarray}
  \label{eq:higgssystem1}
  2 \lambda_0 v^2_0    &=& M^2_{0m} \cos^2\chi + M^2_{1m} \sin^2\chi \\
  \label{eq:higgssystem2}
  2 \lambda_1 v^2_1    &=& M^2_{0m} \sin^2\chi + M^2_{1m} \cos^2\chi \\
  \label{eq:higgssystem3}
  2 \, \eta_1  v_0 v_1 &=&-(M^2_{1m}-M^2_{0m})\, \sin 2\chi\,.
\end{eqnarray}
\\[-5mm]

While the measurement of the two Higgs masses, $M_{0m}$ and $M_{1m}$,
is self-evident, the mixing parameter $\cos\chi$ can be determined
from the Higgs-gauge boson vertex, i.e.
\begin{equation}
   g[H_{0m} W W] = 2 M^2_W \cos\chi /v_0
\end{equation}
where $v_0$ is given by the $W$ mass
\begin{equation}
   M_W = g\, v_0 / 2
\end{equation}
with the SU(2)$_L$ gauge coupling $g$ derived from the measured
$W$-width in a model-independent way to leading order.
[The hypercharge coupling $g'$ is derived correspondingly from combining
the electron electromagnetic-magnetic coupling $e=g s_W$ and with the
hypercharge relation $g'=e/c_W$.]\\

The quartic couplings $\lambda_0, \lambda_1$, or equivalently the
vacuum expectation values $v_0, v_1$, can be separated only by
measuring the triple Higgs couplings. Denoting the current triple
$H_{ic}H_{jc}H_{kc}$ Higgs couplings by $t^c_{ijk}$ $[i,j,k =0. 1]$,
they can be expressed by the physical $H_{pm} H_{qm} H_{rm}$ couplings
$t^m_{pqr}$ $[p,q,r=0,1]$ in the mass basis as
\begin{equation}
  t^c = \, {\cal{O}}^T_H \, \otimes\, {\cal{O}}^T_H \,\otimes
        \, {\cal{O}}^T_H \; t^m   \,.
\end{equation}
The tensor components can be written as
\begin{eqnarray}
  \label{eq:trilinear1}
  t^m_{000} &=& \phantom{+} \tfrac{1}{2}\, M^2_{0m} \, \left(c^3_\chi/v_0
    +s^3_\chi/v_1\right) \\
  \label{eq:trilinear2}
  t^m_{001} &=& -\tfrac{1}{6}\, (2 M^2_{0m}+M^2_{1m})\,
  \left(c_\chi/v_0 -s_\chi/v_1\right)\, c_\chi s_\chi
  \\
  \label{eq:trilinear3}
  t^m_{011} &=& \phantom{+} \tfrac{1}{6}\, ( M^2_{0m}+2 M^2_{1m})\,
  \left(s_\chi/v_0 +c_\chi/v_1\right)\, c_\chi s_\chi
  \\
  \label{eq:trilinear4}
  t^m_{111} &=& - \tfrac{1}{2}\, M^2_{1m} \,
  \left(s^3_\chi/v_0 - c^3_\chi/v_1\right)
\end{eqnarray}
with the abbreviations $c_\chi=\cos\chi$ etc; they are symmetric under
index permutations. The Feynman rules follow from multiplying the
above equations by a minus sign and a combinatorial factor that counts
the number of the identical external legs. The parameter $v_1$ of the
hidden sector is naturally associated either with [small] mixing
coefficients or with coupling/mass suppressed $H_1$ degrees of
freedom. \\

\begin{figure}[!t]
  \parbox{0.44\textwidth}
  {
  \includegraphics[width=8cm]{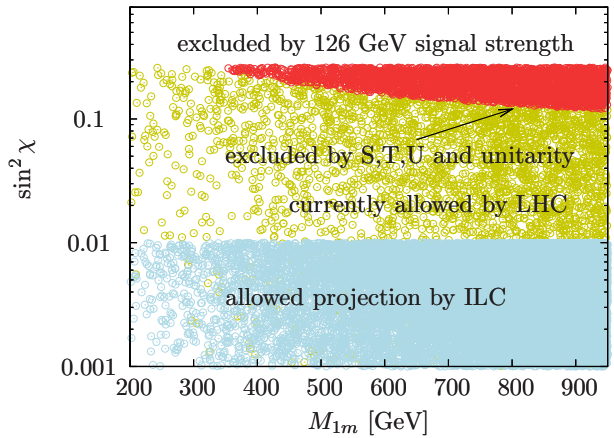}
  }\hspace{0.8cm}
  \parbox{0.44\textwidth}
  {
    \caption{\label{fig:points} Scan over the parameter points of the
      $1\oplus 1$ including current $H_{0m}$ measurements and
      exclusion limits for $H_{1m}$. Kinetic mixing is switched off
      for transparency of the result; invisible Higgs widths are included
      in the scan. Also included are constraints from unitarity and
      oblique corrections \cite{Peskin:1991sw}. We show projections of
      the $M_{1m}$ and $\sin^2\chi$ region that is currently allowed
      by the LHC (yellow) and the parameter region where there will be
      no constraint from a combined ILC+LHC measurement.}
  }
\end{figure}

For illustration purposes, we pick, with $M_{0m} = 125$ GeV, a
representative parameter point
\begin{equation}
  \label{eq:parameter}
  \cos^2\chi = 0.9\,,\quad
  M_{1m}/M_{0m} = 2.5\,,\quad
  v_{1}/v_{0} = 2
\end{equation}
from the scan of the allowed points depicted in
Fig.$\,$\ref{fig:points} [choosing a SM-like width of $H_{0m}$].
Identifying the mass of $H_{0m}$ with 125 GeV, the global area of the
two unknown parameters, i.e. the second Higgs mass $M_{1m}$ and the
mixing $\sin^2\chi$, is tightly constrained by future precision
measurements of the $H_{0m}$ boson. For a numerical investigation of
the above parameter point we adopt the extrapolations to
$3~\text{ab}^{-1}$ for LHC at 14 TeV and adopt an energy of 250 GeV
for ILC as provided in Ref.$\,$\cite{Klute:2013cx}. A measurement of
$\cos^2\chi$ and the masses $M_{1m}$ and $M_{0m}$, which will be well
established at the quoted LHC luminosity, is not enough to separate
vacuum expectation values from quartic couplings in the most general
and complete analysis of the $1\oplus 1 $ system. We need (at least)
one additional measurement in order to reconstruct all the
parameters individually. \\

There are two independent approaches to this problem, depending on the
size of the invisible Higgs branching ratios. For large values, we can
use an invisible Higgs measurement to constrain the parameter
$\sin^2\chi$ as described in Ref.$\,$\cite{NN22}.  \\

Since recent measurements of $H_{0m}$ point towards a SM-like total
width (at least when SU(2)$_L$ doublets are
involved~\cite{Dobrescu:2012td}) we investigate a different
possibility in the following, assuming no direct partial decay width
of $H_{0m}$ to the hidden sector. Then phenomenology is dominated by
mixing only and we can use an experimentally measured region of
$\sin^2\chi$ to constrain the system of mass relations in
Eqs.$\,$\eqref{eq:higgssystem1}, \eqref{eq:higgssystem2} and
\eqref{eq:higgssystem3} as shown in Fig.$\,$\ref{fig:mass}.  The
additional information to reconstruct the individual parameters should
then be made available from the measurement of the trilinear Higgs
couplings in Eqs.$\,$\eqref{eq:trilinear1}-\eqref{eq:trilinear4}
\cite{NN22,Dolan:2012ac}, which can be phenomenologically accessed via
light dihiggs production, i.e. predominantly $gg \to H_{0m}H_{0m}$ at
the LHC \cite{dihiggs}.  [Recall that in $1\oplus n$ scenarios $v_0$
is known from the $W$ mass and the SU(2)$_L$ gauge coupling $g$
measured by the $W$ width, both of which are not affected by U(1)
mixings.] \\

\begin{figure}[!t]
  \parbox{0.44\textwidth}
  {
    \includegraphics[height=5.5cm]{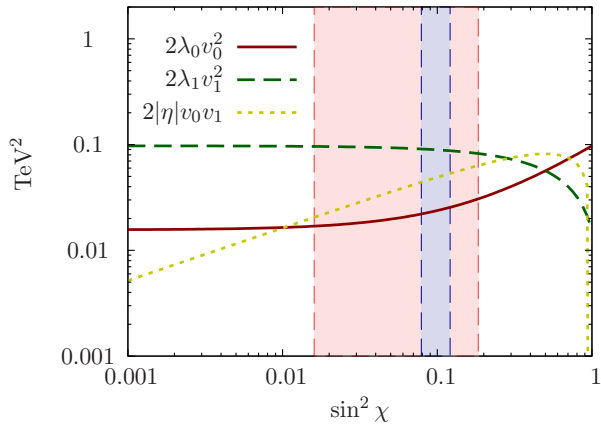}
  }\hspace{1cm}
  \parbox{0.44\textwidth}
  {
    \caption{\label{fig:mass} The system of mass relations in
      Eqs.$\,$\eqref{eq:higgssystem1}, \eqref{eq:higgssystem2} and
      \eqref{eq:higgssystem3} for the parameter point in
      Eq.$\,$\eqref{eq:parameter}. The light red band gives the
      expected $1\sigma$ interval for an LHC measurement of $H_{0m}$
      at the luminosity of $3~\text{ab}^{-1}$. The blue band
      corresponds to the parameter range allowed by an ILC measurement
      at $\sqrt{s}=250$ GeV with the same luminosity of
      $3~\text{ab}^{-1}$.}
  }
\end{figure}

\begin{figure}[!b]
  \setcounter{figure}{2}
  \makeatletter
  \renewcommand{\thefigure}{\@arabic\c@figure a}
  \parbox{0.44\textwidth}
  {
    \includegraphics[height=5.5cm]{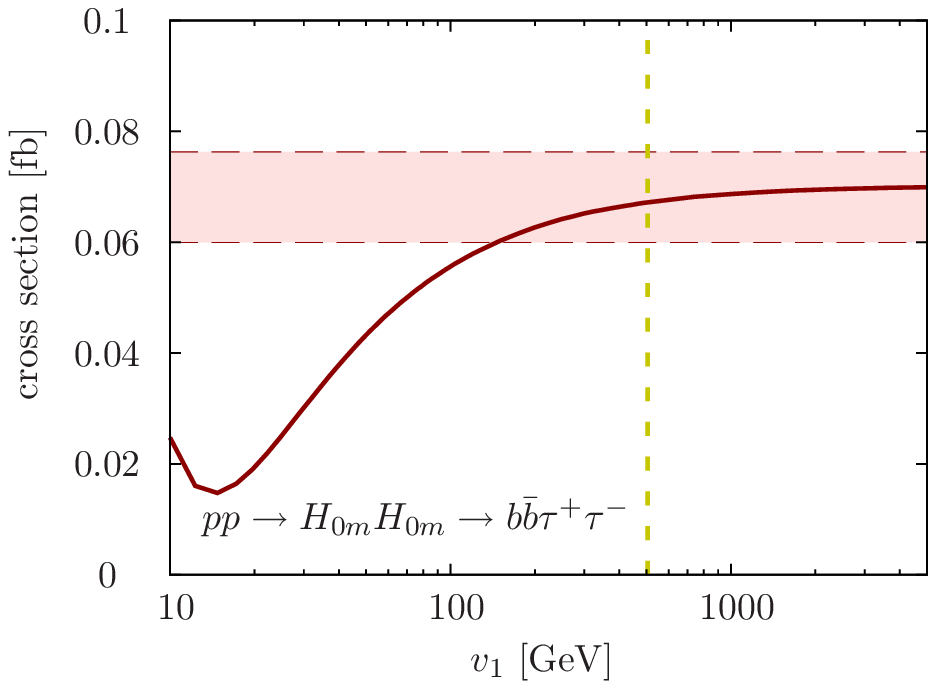}
    \caption{\label{fig:dihiggs1} Dihiggs production cross section for
      the parameter point in Eq.$\,$\eqref{eq:parameter} as function
      of $v_1$ {\it excluding} the $H_{1m}\to H_{0m} H_{0m}$ signal
      region by cutting out the $H_{1m}$ resonance via an invariant
      mass cut on the dihiggs system $m(H_{0m}H_{0m})$.  We use the
      efficiencies of Ref.$\,$\cite{Dolan:2012rv}. The vertical line
      represents the benchmark value of $v_1$ that can be extracted
      from the vector-boson masses in concrete models.}
}
\setcounter{figure}{2}
\makeatletter
\renewcommand{\thefigure}{\@arabic\c@figure b}
\hspace{0.8cm}
  \parbox{0.44\textwidth}
  { \vskip -1\baselineskip
    \includegraphics[height=5.5cm]{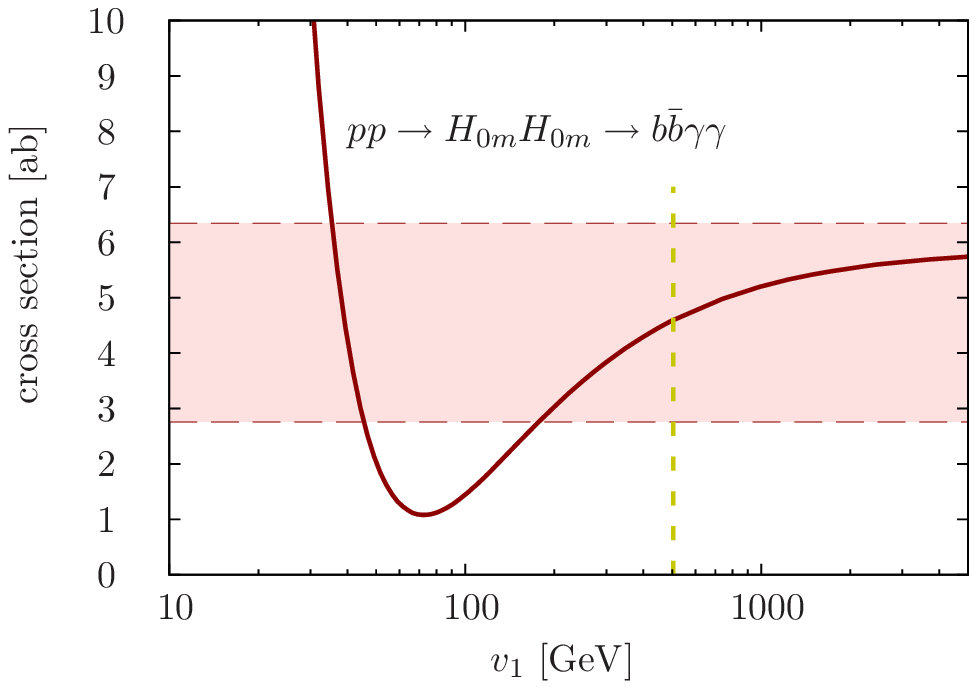}
    \caption{\label{fig:dihiggs2} Dihiggs production cross section for
      Eq.$\,$\eqref{eq:parameter} as function of $v_1$ {\it selecting}
      the $H_{1m}\to H_{0m} H_{0m}$ signal region by cutting out the
      $H_{1m}$ resonance via an invariant mass cut on the dihiggs
      system $m(H_{0m}H_{0m})$. We use the efficiencies of
      Ref.$\,$\cite{ATLAS-collaboration:2012iza}. The vertical line
      represents the benchmark value of $v_1$ that can be extracted
      from the vector-boson masses in concrete models.}
}
\end{figure}
\setcounter{figure}{3}
\makeatletter
\renewcommand{\thefigure}{\@arabic\c@figure}

\begin{figure}[!t]
  \parbox{0.44\textwidth}
  {
    \includegraphics[width=8cm]{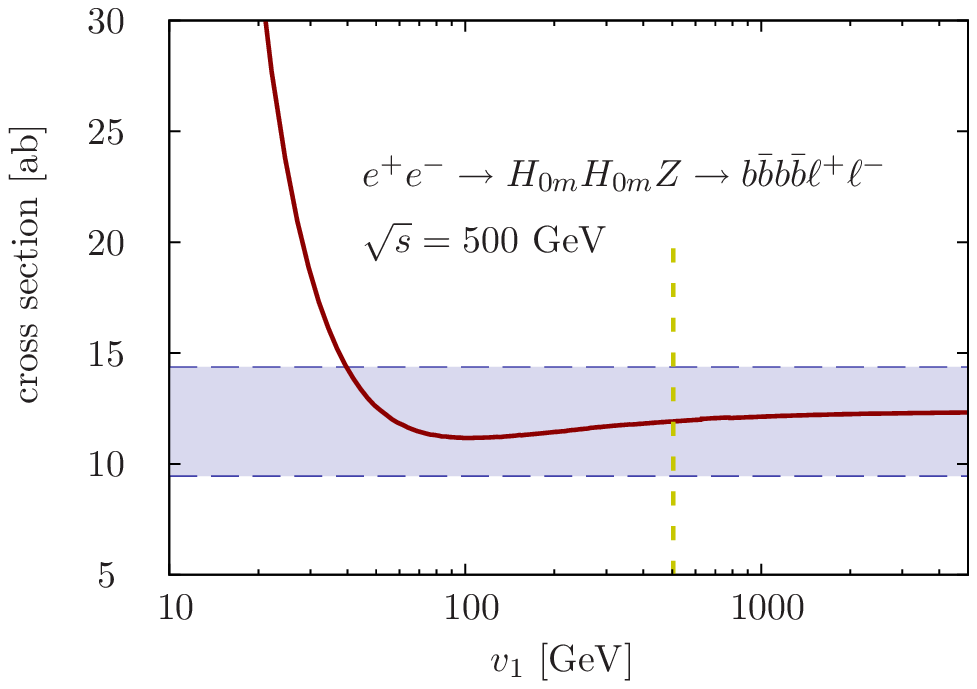}
  }\hspace{0.8cm}
  \parbox{0.44\textwidth}
  {
    \caption{\label{fig:ilc} Double Higgs-strahlung at a 500 GeV
      $e^+e^-$ collider as a function of $v_1$ for the chosen
      parameter point. The blue band corresponds to the parameter
      range allowed by an measurement with a $2\,\text{ab}^{-1}$
      sample. We adopt efficiencies from Ref.$\,$\cite{ilc}.
      }
    }
\end{figure}

Recent analyses \cite{ATLAS-collaboration:2012iza} indicate that a
variation of the trilinear Higgs coupling is only feasible in the
context of the SM in the $b\bar b \tau^+ \tau^-$ channel
\cite{Dolan:2012rv} if possible at all. Rare Higgs decays such as
$H_{0m}H_{0m}\to b\bar b \gamma \gamma$ \cite{Baur:2003gp} are clean
[$S/B=0.7$ for 12 signal events in $3~\text{ab}^{-1}$
\cite{ATLAS-collaboration:2012iza}.] Nevertheless the involved
uncertainties are too large and the signal yield is too small to
obtain a more fine grained picture. In contrast to the SM, however,
the $1\oplus 1$ scenario offers the possibility to discriminate the
$H_{1m}\to H_{0m} H_{0m}$ signal region from the ``continuum''
$H_{0m}H_{0m}$ production. Upon correlating the two regions we can
constrain $v_1$ in different channels: Electroweak precision
measurements, even for rather small mixing angles $\sin^2\chi\sim
0.1$, indicate that the mass splitting between the 125 GeV boson and
$H_{1m}$ must not be too large.  Observing a cascade decay $H_{1m}\to
H_{0m}H_{0m}$ therefore implies small boosts of the $H_{0m}$ bosons
and the analysis of the $H_{0m}H_{0m}\to b \bar b \tau^+\tau^-$ final
state is not applicable anymore. On the other hand, $H_{0m}H_{0m}\to
b\bar b \gamma\gamma$ is inclusive in this sense and we can extract a
limit on $v_1$ in this channel when selecting invariant masses
$m(H_{0m}H_{0m})\sim m(H_{1m})$. The complimentary phase space region
can again be tackled in the boosted selection using the methods of
Ref.$\,$\cite{Dolan:2012rv}. \\

In both analyses interference plays an important role. We therefore
use a complete leading order calculation keeping the full top mass
dependence following Ref.$\,$\cite{Dolan:2012ac}. The result for the
two signal regions including the expected measured $1\sigma$ interval
at 14 TeV and $3~\text{ab}^{-1}$ is shown in
Figs.$\,$\ref{fig:dihiggs1} and~\ref{fig:dihiggs2}. The dip structure
of Fig.$\,$\ref{fig:dihiggs2} highlights the importance of the
interference effects: for $v_1\sim v_0 /\tan\chi \sim 80~{\text{GeV}}$
the resonant production has a global minimum due to the vanishing of
$t^m_{011}$. For values $v_1< v_0/\tan\chi$ the $gg\to H_{1m}\to
H_{0m} H_{0m}$ diagrams interfere destructively with the $gg \to
H_{0m}H_{0m}$ box contributions, but the $t^m_{001}$ grows quickly to
outrun the suppression.  For the away-from-resonance region this
interference is always destructive, i.e. the smaller $v_1$ the larger
$t^m_{000}$ and the smaller the resulting $pp\to H_{0m}H_{0m}$ cross
section until the trilinear coupling $t^m_{001}$ compensates the
enhanced destructive interference of $t^m_{000}$ with the box
contribution and the propagator suppression for very small values of
$v_1$. In this sense $H_{1m}$ ``leaks'' into the $H_{0m}$ measurement
region and must not be discarded
in the actual analysis. \\

The region of large $v_1$ values is determined by the $\cos^2\chi$
pieces of $t^m_{000}$ and $t^m_{001}$ and the asymptotic cross section
settles at a smaller cross section with respect to the SM, mostly as a
consequence of the $\cos^4\chi$ suppression. By contrast, given the
small mixing and the kinematic suppression, it is impossible to
observe $t^m_{011}$ and $t^m_{111}$ at the LHC \cite{Dolan:2012ac}.\\

Depending on the scenario, systematics etc., either the peak or the
continuum analysis can perform better. In any case, both analyses can
be used for cross checks and for lifting the degeneracy, if present,
of the peak analysis. Using the currently known results we find a
lower limit of
\begin{equation}
v_1>200~{\text{GeV}}
\end{equation}
with the $b\bar b \tau^+\tau^-$ analysis to be compared with the
slightly larger bound for the chosen parameter point. This interval
can be mapped onto the allowed region of Fig.$\,$\ref{fig:mass}
constraining $\eta,\lambda_1,\lambda_0$.  However, as can be seen from
Figs.$\,$\ref{fig:dihiggs1} and~\ref{fig:dihiggs2}, the
model-independent separation of the $\lambda$ parameters and the
vacuum expectation value $v_1$ is in general not feasible for the
planned luminosities at LHC as an upper bound on $v_1$ is very loose
if it can be established at all. Even at an $e^+e^-$ collider running
at $\sqrt{s}=500$~GeV (Fig.$\,$\ref{fig:ilc}) we can only extract a
lower limit on $v_1$ at an integrated luminosity of $2~\text{ab}^{-1}$
[see Ref.~\cite{Djouadi:1999gv} for a general discussion of measuring
the trilinear Higgs coupling at a linear collider].  However, given
that at this luminosity the uncertainties are still statistics-driven,
there might be the possibility to extract an upper limit on $v_1$ in
the far future. In fact, the quoted uncertainty band is entirely
dominated by the statistical uncertainty of the signal counts as the
search is
essentially background-free \cite{ilc}. \\

However, in theoretical scenarios which predict the values of the
gauge coupling and the hypercharge in the hidden sector, the vacuum
expectation value $v_1$ can be determined from the two vector-boson
masses:
\begin{equation}
   v_1 / v_0 = [g/2]/[g_V Y_V]\,\times\, M_{Vc}/ M_W   \,.
\end{equation}
\\[-5mm]

Thus, the fundamental current parameters $\lambda_0,\mu^2_0;
\lambda_1, \mu^2_1; \eta$ in the Higgs potential can, in principle, be
extracted from experimental data, the combinations $\lambda_0 v^2_0;
\lambda_1 v^2_1; \eta_1 v_0 v_1$ easily extracted from masses and
mixings, and the $v'$s separately bounded from trilinear Higgs
couplings or derived from vector-mass measurements in specified
theories. When the gauge couplings and charges are predicted
theoretically, all the fundamental Higgs parameters can be
extracted.\\

\subsection{Kinetic Mixing}

\noindent
Analogously, the mixing of the gauge sector can be worked out
explicitly.  The kinetic term and the mass term are diagonalized
by a ${\rm SL}(3,{\rm R})$ kinetic transformation and an orthogonal
$3\times 3$ rotation matrix ${\cal{O}}_V$ as
\begin{eqnarray}
\left( \begin{array}{c}
       W \\
       B \\
       V
       \end{array}\right)_c
=
\left( \begin{array}{ccc}
       s_W    &   c_W   &           0 \\[2mm]
       c_W    &  -s_W   & - s \sigma  \\[2mm]
       0      &   0     &   \sigma
       \end{array}\right)
\left( \begin{array}{ccr}
         1         &  0            &          0  \\[2mm]
         0         &   \cos\theta  & -\sin\theta \\[2mm]
         0         &   \sin\theta  & \cos\theta
       \end{array}\right)
\left( \begin{array}{c}
        A \\
        Z \\
        V
       \end{array}\right)_m
\end{eqnarray}
with $\sigma = 1/\sqrt{1-s^2}$. The masses, in the current basis,
\begin{eqnarray}
   M^2_{W_c} & = & g^2 \, T^2_3\, v^2_0 \, = g^2\, v^2_0 /4    \\
   M^2_{B_c} & = & g'^2\, Y^2_0\, v^2_0 = g'^2 v^2_0/4         \\
   M^2_{Z_c} & = & [g^2+g'^2] v^2_0/4                          \\
   M^2_{V_c} & = & g^2_V\, Y^2_V\, v^2_1
\end{eqnarray}
are defined by the gauge couplings, the vacuum expectation values, and
the SU(2) $T_3$ and the U(1) $Y$ charges of the Higgs fields. Since
the charged $W$-field does not mix with the vector field in the hidden
sector, the measured values of the $W$-mass and width determine the
parameters $g$ and $v_0$ and the SM relations $g = e/s_W$ and $g'=
e/c_W$ define $s_W$, $g'$, $g_Z = [g^2+g'^2]^{1/2}$ and $M_{Z_c}$
before mixing. The [neutral] current masses are transformed to the
vanishing photon mass $M_{A_m}=0$ and two physical non-zero gauge
boson masses $M_{Z_m}, M_{V_m}$ by the rotation angle~$\theta$. The
exact and the approximate forms, expanded up to second order in $s$
and $||M^2_{Zc}/M^2_{V_C}||$, may be denoted as
%
\begin{eqnarray}
M^2_{Z_m} & = & M^2_{Z_c}
               \left\{ (1+s^2_W \sigma^2 s^2 + \sigma^2 \Delta )
               - \sqrt{(1+s^2_W \sigma^2 s^2 +\sigma^2 \Delta)^2
                        - 4 \sigma^2 \Delta}\right\}/2 \simeq M^2_{Z_c} + ...
               \label{eq:mBm_squared}       \\
M^2_{V_m} & = & M^2_{Z_c}
               \left\{ (1+s^2_W \sigma^2 s^2 + \sigma^2 \Delta )
               + \sqrt{(1+s^2_W \sigma^2 s^2 + \sigma^2 \Delta)^2
                        - 4 \sigma^2 \Delta}\right\}/2
               \,\simeq\, M^2_{Vc}+ s^2 M^2_{V_c} + ...
               \label{eq:mVm_squared} \\
\tan 2\theta &=& 2 s_W \sigma s/(1-s^2_W \sigma^2 s^2-\sigma^2\Delta)
               \,\simeq\, -2 s_W (M^2_{Z_c}/M^2_{V_c})\, s + ...
               \label{eq:tan2theta}
\end{eqnarray}
with $\Delta = M^2_{V_c}/M^2_{Z_c}$. \\

The vertices relevant for measuring the parameters of the theory, are
collected in Tab.~\ref{tab1}.\\

\begin{table}
\begin{tabular}{|l||l|l|l|}
\hline
\ \ vertex                 &  \ \     type
                        &  \ \  exact
                        &  \ \  expansion in mixing       \\
\hline\hline
\ \ 3-gauge bosons         & \ \ $A_m WW$
                        & \ \ $e$
                        & \ \ $e$                         \\[0.2mm]
                       & \ \ $Z_m WW$
                        & \ \ $g_Z c^2_W c_\theta$
                        & \ \ $g_Z c^2_W$                 \\[0.2mm]
                       & \ \ $V_m WW$
                        & \ \ $-g_Zc^2_W s_\theta$
                        & \ \ $ g_Z c^2_W s_W s\, M^2_{Z_c}/M^2_{V_c}$   \\[1mm]
\hline
\ \ gauge boson - fermion\ \
     & \ \ $A_m ff$
      &  \ \ $ eQ_f$
      &  \ \ $ e Q_f$                         \\[0.2mm]
     & \ \ $Z_m ff$
      & \ \ $g_Z (T^f_3-s^2_W Q_f) c_\theta
              -g' Y_f\, \sigma s\, s_\theta$
      & \ \ $g_Z (T^f_3-s^2_W Q_f) $          \\[0.2mm]
     & \ \ $V_m ff$
      &  \ \ $-g_Z (T^f_3-s^2_W Q_f) s_\theta
              -g' Y_f\, \sigma s\, c_\theta$
      &  \ \ $ - g' Y_f\, s$                  \\
     &  &
      & \ \  $+ g_Z (T^f_3-s^2_W Q_f)s_W s M^2_{Z_c}/M^2_{V_c}$    \\[1mm]
\hline
\ \ Higgs - gauge boson
     & \ \ $H_{0m} WW$   \ \
      & \ \ $2 M^2_{W_c} c_\chi/v_0$
      & \ \ $2 M^2_{W_c}[1- \frac{1}{4}(\eta_1 v_0/\lambda_1 v_1)^2]/v_0$
        \\[0.2mm]
     & \ \ $H_{1m} WW$  \ \
      & \ \ $-2 M^2_{W_c} s_\chi/v_0$
      & \ \ $M^2_{W_c} (\eta_1/\lambda_1)/ v_1$  \\[0.2mm]
     & \ \ $H_{0m} Z_m Z_m$  \ \
      & \ \ $ 2 M^2_{Z_c}
             [(c_\theta+ s_W \sigma s s_\theta)^2
             (c_\chi/v_0)
             + \sigma^2\Delta\, s^2_\theta\,
             (s_\chi/v_1)]$\ \
      &  \ \ $2 M^2_{Z_c}[1- \frac{1}{4}(\eta_1 v_0/\lambda_1 v_1)^2]/v_0$
        \\[0.2mm]
     & \ \ $H_{1m} Z_m Z_m$ \ \
      & \ \ $ 2 M^2_{Z_c} [(c_\theta+ s_W \sigma s s_\theta)^2
                           (-s_\chi/v_0)
                           + \sigma^2\Delta\, s^2_\theta\,
                           (c_\chi/v_1)]$\ \
       & \ \ $M^2_{Z_c} (\eta_1/\lambda_1)/v_1$    \\[0.2mm]
      & \ \ $H_{0m} Z_m V_m$ \ \
       & \ \  $2 M^2_{Z_c}
              [(c_\theta+ s_W \sigma s s_\theta)
              (-s_\theta+s_W \sigma s c_\theta)
              (c_\chi/v_0)$ \ \
       & \ \ $ 2 M^2_{Z_c} s_W s [1- M^2_{Z_c}/M^2_{V_c}] /v_0$    \\[0.2mm]
       &
       & \ \ \hskip 1.cm
             $+ \sigma^2\Delta\, c_\theta s_\theta\,
                                 (s_\chi/v_1)]$
       &                                                \\[0.2mm]
      & \ \ $H_{1m} Z_m V_m$ \ \
       & \ \  $2 M^2_{Z_c}
             [(c_\theta+ s_W \sigma s s_\theta)
             (-s_\theta+s_W \sigma s c_\theta)
             (-s_\chi/v_0)$ \ \
       & \ \ $ M^2_{Z_c} s_W
               [2+\eta_1/\lambda_1]\,s/v_1$  \ \         \\ [0.2mm]
       &
       & \ \ \hskip 1.cm
       $+ \sigma^2\Delta\, c_\theta s_\theta\, (c_\chi/v_1)]$
       &                           \\[1mm]
\hline
\end{tabular}
\caption{\label{tab1} 3-vertices of gauge bosons, fermions and Higgs bosons.
  The vertices are expanded up to non-vanishing first/second order in the
  kinetic and Higgs mixings. [Standard tensor and fermion bases are not
  noted explicitly.]}
\end{table}

\begin{figure}[!b]
  \setcounter{figure}{4}
  \makeatletter
  \renewcommand{\thefigure}{\@arabic\c@figure a}
  \parbox{0.44\textwidth}
  {  \vskip -1\baselineskip
    \includegraphics[height=5.5cm]{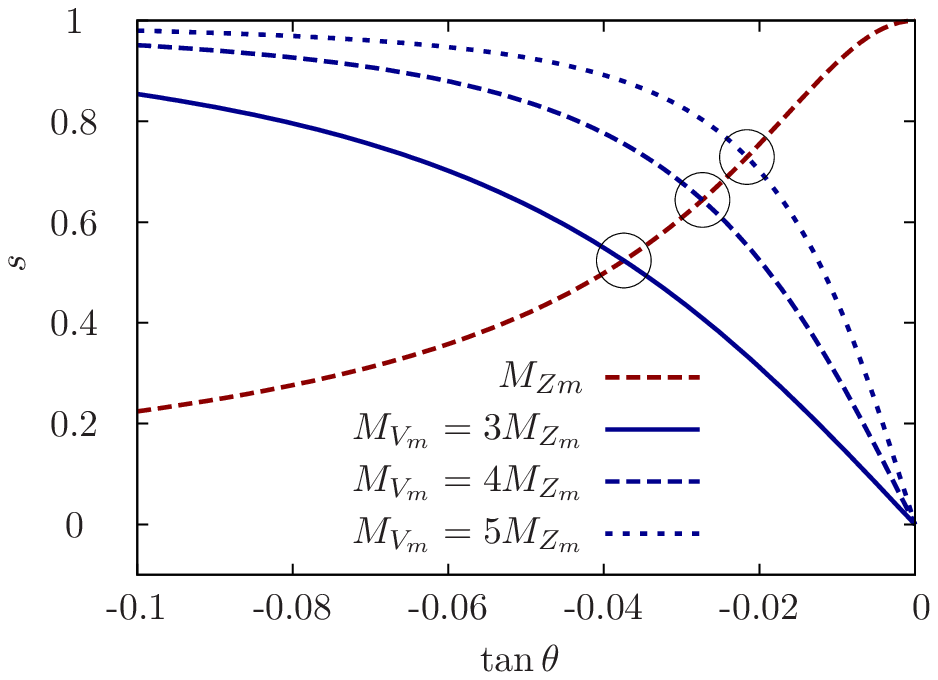}
    \caption{\label{fig:theta} Extraction of the kinetic mixing
      parameters $s$ and $\tan\theta$ from a sample of measured values
      $M_{Z_m}$ and $M_{V_m}$ from Eqs.$\,$\eqref{eq:vecmasses11} and
      \eqref{eq:vecmasses12}.  }
  }\hspace{0.8cm}
  \setcounter{figure}{4}
  \makeatletter
  \renewcommand{\thefigure}{\@arabic\c@figure b}
  \parbox{0.44\textwidth}
  {
    \includegraphics[height=5.5cm]{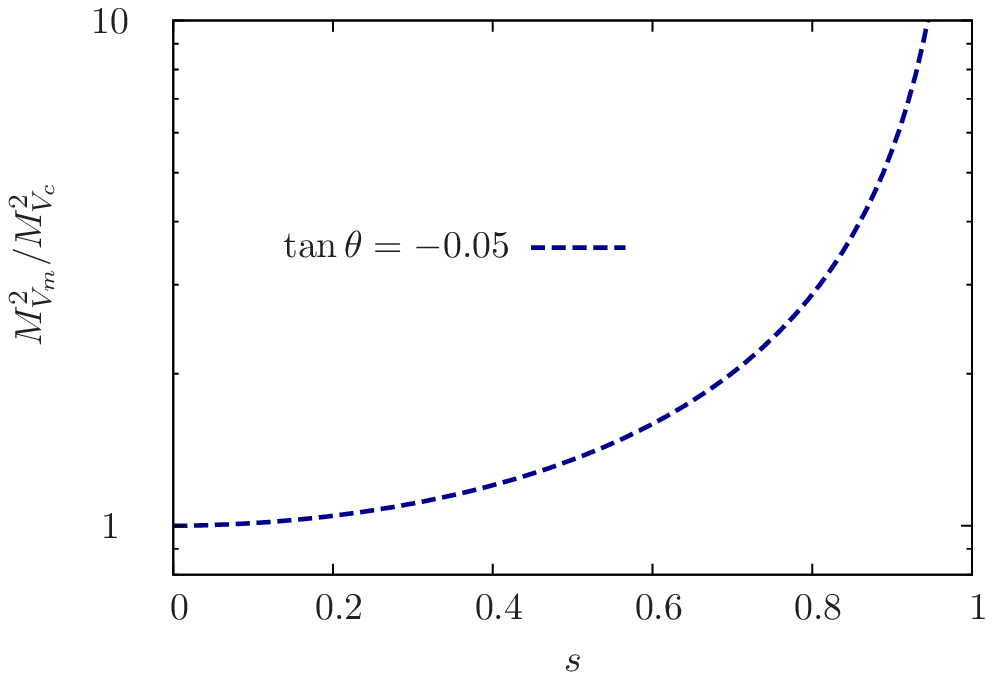}
    \caption{\label{fig:s} Shift of the mass $M_{V_m}$ from $M_{V_c}$
      in the hidden sector, depending on the kinetic mixing parameter
      $s$ for a representative rotation parameter $\tan\theta$, from
      Eq.$\,$\eqref{eq:vecmasses2}.}
  }
\end{figure}

The direct measurement of the $W, Z_m, V_m$ mass parameters [see
e.g.~Refs.$\,$\cite{zprimea,zprimec} for $V_m$ searches as performed
by ATLAS and CMS] in the mass basis together with the indirect
determination of the mass parameters in the current basis in
Eqs.$\,$\eqref{eq:mBm_squared}, \eqref{eq:mVm_squared} and
\eqref{eq:tan2theta} allows for a complete analysis of the fundamental
vector parameters, including the KT parameter $s$ and the rotation
angle $\theta$. From the two relations
\begin{eqnarray}
  \label{eq:vecmasses11}
  M^2_{Z_m} /  M^2_{Z_c} &=& 1 + s_W \frac{s}{\sqrt{1-s^2}}\, \tan\theta \\
  \label{eq:vecmasses12}
  M^2_{V_m} /  M^2_{Z_c} &=& 1 - s_W \frac{s}{\sqrt{1-s^2}}\, \cot\theta
\end{eqnarray}
the mixing parameters, $s$ and $\tan\theta$ can be extracted,
Fig.$\,$\ref{fig:theta}. [Note that due to pure U(1) mixing the weak
couplings and $v_0$ are fixed and measurable quantities, as outlined
above, and $M_{Z_c}$ is known.] These values can be exploited to
derive the third mass parameter $M_{V_c}$ in the current basis:
\begin{eqnarray}
  \label{eq:vecmasses2}
 M^2_{V_c} /  M^2_{V_m} &=& (1-s ^2)\, M^2_{Z_m}/M^2_{Z_c} \nonumber\\
  &=& 1 - s^2 + s_W s \sqrt{1-s^2} \tan\theta
\end{eqnarray}
cf. Fig.$\,$\ref{fig:s}.  These results can be cross-checked for
internal consistency by experimentally analyzing vertices collected in
Tab.$\,$\ref{tab1}. Note that oblique corrections due to kinetic
mixing are typically less important as compared to Higgs mixing since
they scale $\sim s^2M_{Z_m}^2/M_{V_m}^2$
\cite{Holdom:1990xp}. Explicit calculation proves that they lead to
bounds which are comparable to the limits currently set by ATLAS and
CMS (see below).\\

Current constraints on anomalous triple gauge boson vertices, obtained
by ATLAS \cite{tripa} and CMS \cite{tripc}, restrict $\sin^2\theta$ to
$\lesssim 0.1$, Fig.$\,$\ref{fig:ww1}, which becomes comparable to the
LEP combination \cite{Schael:2013ita}. The $WW$ analysis is highly
sensitive to modifications denoted in Tab.$\,$\ref{tab1}, in light of
a tree-level radiation zero in the SM \cite{Brown:1982xx}. There
exists a completely destructive interference between the $W$ radiation
diagrams and the Feynman graph involving the $ZWW$ vertex in the
SM. Any (non-global) deviation from the SM-predicted coupling pattern
destroys this characteristic angular dip structure resulting in an
increase of the total cross section.\footnote{This is strictly true at
  tree level. Gluon induced channels of the higher order-corrected
  cross section lift the radiation zero.}  Therefore, the integrated
cross section [supplemented by a jet veto in the actual analysis] is
very sensitive to the suppression of the $Z_m WW$ coupling, which can
be used to extract an upper limit on $\sin^2\theta$ even if the $V_m$
is too heavy to be measured directly. \\

%
\begin{figure}[!t]
  \setcounter{figure}{5}
  \makeatletter
  \renewcommand{\thefigure}{\@arabic\c@figure}
  \parbox{0.44\textwidth}
  {
    \includegraphics[width=8cm]{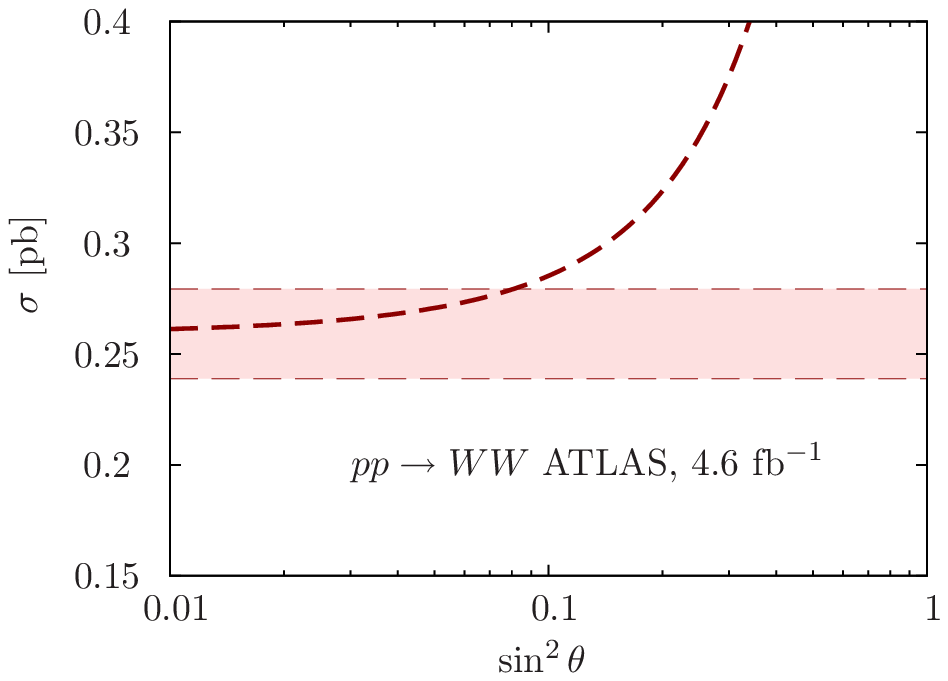}
  }\hspace{0.8cm}
  \parbox{0.44\textwidth}
  {
    \caption{\label{fig:ww1} $pp\to WW$ cross section as reported in
      Ref.$\,$\cite{tripa} and the cross section as a function of
      $\sin^2\theta$ for $s$ neglected after removing the $V_m$
      resonance contribution.}
  }
\end{figure}

\begin{figure}[!b]
  \parbox{0.44\textwidth}
  {
  \includegraphics[width=8cm]{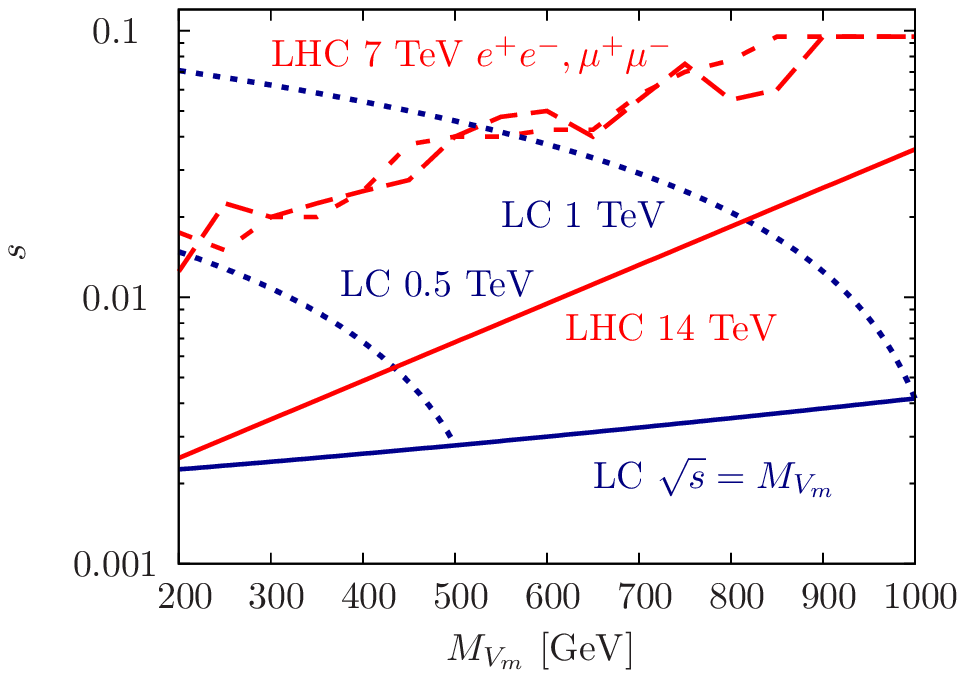}
  }\hspace{0.8cm}
  \parbox{0.44\textwidth}
  {
    \caption{\label{fig:kinmix} 95\% exclusion limits in the
      $[M_{V_m},s]$ plane for the 7 TeV ATLAS $V_m$ analysis (for
      $V_m\to \mu^+\mu^-$ and $V_m\to e^+ e^-$
      \cite{zprimea,zprimec,Jaeckel:2012yz}) and a projection based on
      the extrapolation of these results to 14 TeV and 300
      fb$^{-1}$. The LC limit corresponds to $e^+e^-\to
      {\text{visible}}$ excluding top quarks and Bhabha scattering. We
      have chosen center-of-mass energies of 500 GeV and 1 TeV (dotted
      curves) with a bremsstrahlung spectrum modelled as reported in
      Ref.$\,$\cite{boor}; the full line predicts the maximum sensitivity
      to the mixing parameter $s$ for a scan at the c.m. energy
      $\sqrt{s} \simeq M_{V_m}$. Note, that these exclusion limits are
      sensitive to the dynamics in the hidden sector via the $V_m$
      branching ratio; we only take partial decay widths to SM matter
      into account in this plot.}
  }
\end{figure}

Note that the $H_{0m}WW$ coupling is not affected by gauge-kinetic
mixing so that it can be used to measure $\cos\chi$ individually. \\

Subsequently, the measurement of the $H_{0m} Z_mZ_m$ vertex can in
principle be used to measure the allowed range of the shift of
Tab.$\,$\ref{tab1}.  Some concrete models of U(1) mixing in the field
and string theories \cite{kinmix,Hportal_kinmix} predict the mixing
parameter to be $\lesssim 10^{-3}$, but string models in general allow
for rather large $s$ values well in the sensitivity range of LHC (and
ILC).  A hierarchy $s \ll \eta$ then results in modification of the
$H_{0m}$ couplings dominated by the Higgs mixing.  This leaves
Drell-Yan--like production of $V_m$ as probably the best search
strategy in the mass region that can be covered by the LHC and future
colliders, Fig.$\,$\ref{fig:kinmix}. Once the vector-boson mass range
is known from LHC, LC scans can improve the sensitivity to
measurements of the kinetic mixing parameter $s$.\\


\section{$1\oplus 2$ ANALYSIS}

\noindent
To exemplify the analytic and numerical analyses of the preceding
sections further, we shall discuss next a more complicated system
where the SM Higgs sector is connected by portal interactions with two
Higgs systems in the hidden sector, and the U(1) hypercharge field $B$
is mixed kinematically with two abelian U(1) fields $V_{1,2}$ in the
hidden sector with charges $Y_{V_{1,2}}$. Such a system may be
generated quite easily in superstring theories. \\

The microscopic parameters in the Higgs sector are chosen such that
physical masses are centered around 125, 300, 1000 GeV before mixing.
For illustration purposes we take these mass parameters to derive
$v_1,v_2$ for quartic Higgs-coupling values
$\lambda_0=\lambda_1=\lambda_2$ and set $\eta_2=\eta_1/2$, such that
$\eta=\eta_1$ is the only free parameter left. The $B$ mass associated
with the Standard Model is fixed at $M_B = \tfrac{1}{4} g'^2 v^2_0$, the
gauge masses in the hidden sector are chosen as 300 and 1000 GeV.  The
kinetic mixing vector in the gauge sector, defined as $s = |s|
[\cos\phi,\sin\phi]$, is varied by running $|s|$ from 0 to 1, with
$\sin\phi = 1/\sqrt{2}$ kept fixed. This choice of free parameters is
chosen minimal for the sake of transparency. \\

\subsection{Higgs system}

\noindent
Reading off the bilinear mass terms from the Higgs-portal Lagrangian,
the Higgs mass matrix turns out to be
\begin{eqnarray}
  \label{eq:12eval}
  \mathcal{M}^2_{Hc}
= \left( \begin{array}{ccc}
         2\lambda_0 v^2_0  & \eta_1 v_0 v_1   & \eta_2 v_0 v_2   \\
         \eta_1 v_0 v_1    & 2\lambda_1 v^2_1 & 0                \\
         \eta_2 v_0 v_2    & 0                & 2\lambda_2 v^2_2
         \end{array}  \right)
         \Rightarrow
  \left( \begin{array}{cc}
         M_{0c}^2   & X^T    \\
         X       & M^2_c
         \end{array}  \right) \,.
\end{eqnarray}
This $3\times 3$ matrix could in principle be diagonalized
analytically, but the result is not transparent anymore. In the case
of small mixing the approximate results of the previous section can be
applied. In general, however, numerical methods will be adopted and
the results will be illustrated in figures. \\

The three eigenvalues of the mass matrix ${\mathcal{M}}^2_{Hc}$ are
displayed in Fig.$\,$\ref{fig:12mass} for the fixed parameters as defined
earlier, but the mixing parameter $\eta$ parameterizing the value of
$X$ is varied. For small mixing, and large scales of the hidden sector,
the masses are given analytically by
\begin{eqnarray}
M^2_{0m} &=& 2 \lambda_0 v^2_0
            - \eta_1^2 v_0^2/2\lambda_1
            - \eta_2^2 v_0^2/2\lambda_2              \\
M^2_{1m} &=& 2 \lambda_1 v^2_1
            + \eta_1^2 v_0^2/2\lambda_1              \\
M^2_{2m} &=& 2 \lambda_2 v^2_2
            + \eta_2^2 v_0^2/2\lambda_2
\end{eqnarray}
in second order approximation of the mixing.\\

The orthogonal matrix $\mathcal{O}_H$ in Eq.$\,$\eqref{eq:O_H} is most
elegantly parameterized by the three rotation angles
$\theta_{01},\theta_{02},\theta_{12}$ associated with the rotations of
the $01,02,12$ planes, i.e.  $\mathcal{O}_H=R_2(\theta_{01})R_1^T
(\theta_{02}) R_0(\theta_{12})$.  The sine-functions of the angles are
shown in Fig.$\,$\ref{fig:12angle} for varied mixings $X$ as defined
above. For small mixing the sines of the rotation angles read
\begin{eqnarray}
\sin\theta_{01} &\simeq& - \eta_1 v_0/2\lambda_1 v_2 \\
\sin\theta_{02} &\simeq& - \eta_2 v_0/2\lambda_2 v_2 \\
\sin\theta_{12} &\simeq& 0
\end{eqnarray}
up to linear order in the mixing.\\

\begin{figure}[!t]
  \setcounter{figure}{6}
  \makeatletter
  \renewcommand{\thefigure}{\@arabic\c@figure a}
  \parbox{0.44\textwidth}
  { \vskip -1\baselineskip
    \includegraphics[height=5.5cm]{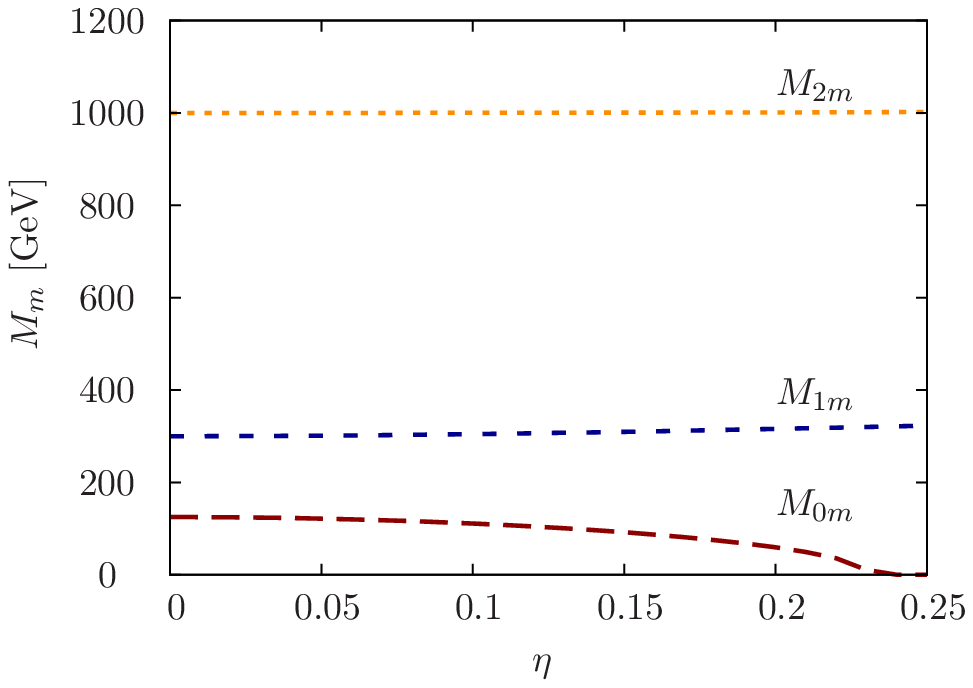}
    \caption{\label{fig:12mass} Eigenvalues of the matrix
      Eq.$\,$\eqref{eq:12eval} as a function of the
      Higgs mixing. The choice of parameters before mixing is
        described in the text.}
    }\hspace{0.8cm} \setcounter{figure}{6} \makeatletter
    \renewcommand{\thefigure}{\@arabic\c@figure b}
  \parbox{0.44\textwidth}
  {
    \includegraphics[height=5.5cm]{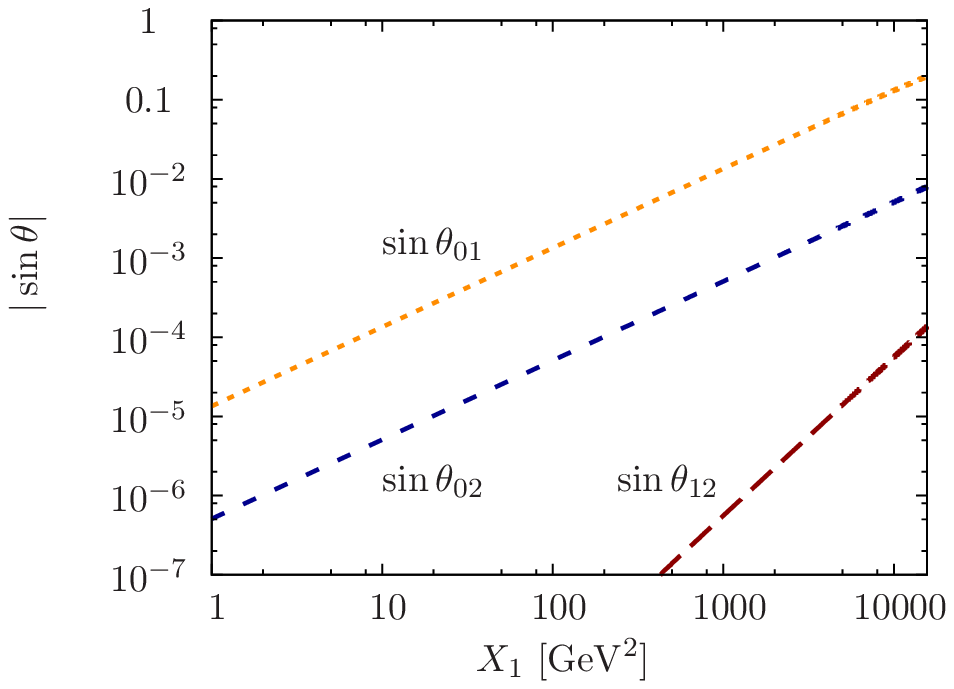}
    \caption{\label{fig:12angle} Sines of the rotation angles that
      diagonalize the mass matrix in Eq.$\,$\eqref{eq:12eval} via the
      orthogonal matrix $\mathcal{O}_H=R_2(\theta_{01})R_1^T
      (\theta_{02}) R_0(\theta_{12})$.  The parameter choices are
      described in the text and are identical with
      Fig.$\,$\ref{fig:12mass}.  }
  }
\end{figure}
  \setcounter{figure}{6}
  \makeatletter
  \renewcommand{\thefigure}{\@arabic\c@figure}

Given the parameters, masses and mixings, of the physical states,
the production cross sections and the decay branching ratios can be
predicted in the standard way. \\

The mixing modifies the couplings of the original SM Higgs boson to
the $W$-gauge bosons and it generates the corresponding couplings to
the original Higgs bosons in the hidden sector:
\begin{equation}
   \{ H_{0m}; H_{1m}; H_{2m} \}\, WW
= 2 M^2_{W_c} / v_0 \times \{ \cos{\theta_{01}} \cos{\theta_{02}} ;
                              -\sin{\theta_{01}} \cos{\theta_{02}} ;
                              -\sin{\theta_{02}} \}   \,.
\end{equation}
The coupling to $Z$-boson pairs is more involved if the kinetic
mixing modifies the current $Z$ eigenstates to mass eigenstates
as discussed later. \\

In the same way as above, the triple Higgs couplings in the potential
can be derived from the triple Higgs couplings of the mass eigenfields
which are accessible, in principle, from experimental data of multiple
Higgs production cross sections; in self-explaining notation for small
mixing:
\begin{eqnarray}
   t^m_{000} &\simeq&  \frac{M^2_{0m}}{2 v_0}\, \cos^3{\theta_{01}}
                       \cos^3{\theta_{02}}                               \\
   t^m_{001} &\simeq&  \frac{(2 M^2_{0m}+M^2_{1m})}{6}
                       \left(\frac{\sin\theta_{01}}{v_1}-\frac{1}{v_0}\right)
                       \quad {\rm and} \quad  \{ 1 \leftrightarrow 2 \}  \\
   t^m_{011} &\simeq&  \frac{(M^2_{0m}+2 M^2_{1m})}{6}
                       \left(\frac{\sin\theta_{01}}{v_0}+\frac{1}{v_1}\right)
                       \quad {\rm and} \quad  \{ 1 \leftrightarrow 2 \}  \\
   t^m_{012} &\simeq&  \frac{(M^2_{0m}+M^2_{1m}+M^2_{2m})}{6 v_0}\,
                       \sin\theta_{01} \sin\theta_{02}                   \\
   t^m_{111} &\simeq&  \frac{M^2_{1m}}{2 v_1} \cos^3\theta_{01}
                       \quad {\rm and} \quad  \{ 1 \leftrightarrow 2 \}  \\
   t^m_{112} &\simeq&  \frac{M^2_{0m} (2M^2_{1m}+M^2_{2m})}{
                             6(M^2_{1m}-M^2_{2m})\,v_1}\,
                             \sin\theta_{01} \sin\theta_{02}
                       \quad {\rm and} \quad  \{ 1 \leftrightarrow 2 \}
\end{eqnarray}
symmetric under permutations of the Higgs indices. \\

Thus, three Higgs masses, the couplings of the Higgs bosons to SM pairs
and the triple Higgs couplings allow the determination of the current
parameters in the Higgs potential. \\

\subsection{Kinetic Mixing}

The mixing in the gauge sector can be illustrated in a similar
fashion. Disregarding potential mixing of the two abelian gauge fields
$V_{1,2}$ within the hidden sector [which could be implemented with no
problem], the kinetic mixing of the hypercharge $B$-field with the
abelian gauge fields, and its weak mixing with the neutral SU(2)
$W$-field, affect masses and couplings of the mass vector-eigenstates. \\

The $2 \times 2$ KT matrix $\sigma$ introduced in
Eq.$\,$(\ref{eq:stretch}) is diagonalized by the orthogonal matrix
\begin{equation}
   u = \left( \begin{array}{rr}
       \cos\phi   & \sin\phi  \\
       - \sin\phi & \cos\phi
       \end{array} \right)
     = \frac{1}{|s|}
       \left( \begin{array}{rr}
              s_1 & s_2 \\
             -s_2 & s_1
              \end{array} \right)
\end{equation}
with $|s|=(s^2_1+s^2_2)^{1/2}$ for $s = (s_1,s_2)$. The KT matrix
$\sigma$ turns into a diagonal matrix ${\rm diag}\,[\sigma^\prime,\,
1]$ with $\sigma^\prime=(1-|s|^2)^{-\frac{1}{2}}$. The $\mathcal{Z}$
matrix, transforming the current gauge vector fields $\{W, B, V_1,
V_2\}_c$ to the mass vector fields $\{A,Z, V_1, V_2\}_m$ is given by
\begin{eqnarray}
  \mathcal{Z}
= \left( \begin{array}{cccc}
           s_W &  c_W  & 0 & 0 \\[2mm]
           c_W & -s_W  & 0
                       & 0   \\[2mm]
           0   & -(\sigma^{\prime 2}-1)^{1/2} c_\phi\,
               &  \sigma^\prime c^2_\phi+ s^2_\phi\,
               &  \,(\sigma^\prime-1) c_\phi s_\phi\,    \\[1mm]
           0   & -(\sigma^{\prime 2}-1)^{1/2} s_\phi
               & \,(\sigma^\prime-1) c_\phi s_\phi\,
              &  \,\sigma^\prime s^2_\phi+ c^2_\phi\,
         \end{array}  \right)
\simeq
\left( \begin{array}{cccc}
           s_W  &  c_W  &  0  & 0          \\[2mm]
           c_W  &  -s_W &  0  & 0          \\[2mm]
           0    & -s_1  &  \, 1+ s^2_1/2\,
                &   s_1 s_2/2              \\[1mm]
           0    & -s_2  &  \, s_1 s_2/2\,
                &   1 + s^2_2/2
         \end{array}  \right)
\end{eqnarray}
up to the second order of the mixing $|s|$, with the abbreviations
$c_\phi=\cos\phi=s_1/|s|$, etc. The $\mathcal{Z}$ matrix affects the
vector-boson masses and the charges when varying the KT factor
$\sigma^\prime$ and the angle $\phi$ in the mixing column vector $s$.\\

For a small mixing parameter $|s|\ll 1$, the three massive gauge boson
masses, in addition to a vanishing photon mass $M_{A_m}=0$, are given
approximately by
\begin{eqnarray}
M^2_{Z_m}  &=& M^2_{Z_c}  \\
M^2_{V1_m} &=& M^2_{V1_c} +M^2_{V1_c} s^2_1 \\
M^2_{V2_m} &=& M^2_{V2_c} +M^2_{V2_c} s^2_2
\end{eqnarray}
up to the second order in the mixing. By construction, the isospin
$T_3$ and hypercharge $Y$ are not changed {\it a priori} to leading
order, while the generalized $V$ charges can be derived from the
transformed covariant derivative
\begin{eqnarray}
   T_3\, W_c &\Rightarrow & s_W\, T_3\, A_m + c_W T_3\, Z_m     \\[1.5mm]
   Y B_c     &\Rightarrow & c_W Y A_m - s_W\, Y Z_m             \\[1.5mm]
   Y_{V1} V_{1c}+ Y_{V2} V_{2c} &\Rightarrow&
   (Y_{V1} - [g'/g_V] Y s_1 )\, V_{1m} + ( Y_{V2} - [g'/g_V] Y s_2 )\, V_{2m}
\end{eqnarray}
to leading order in the mixing. \\

The exact mass eigenvalues of $V_{m}$ are illustrated in
Fig.$\,$\ref{fig:charge} as a function of the kinetic mixing
parameters $|s|$ and $\cos\phi = 1/\sqrt{2}$.  The charge components
correspondingly in Fig.$\,$\ref{fig:charge} for couplings $g_V /
g^\prime$ and $Y_V / Y \Rightarrow 1$. \\

The $|s|$-dependence in these figures is easy to understand. For large
enough $|s|\to 1$ the $4\times 4$ mass matrix approaches the limit of
two $2\times 2$ system with only weak coupling. The masses in the hidden
sector grow for $|s|\to 1$, leading to strong $|s|$ dependence of the
heavier hidden mass eigenvalue as a consequence of the diagonalization
of the hidden $2\times 2 $ symmetric submatrix. The coefficient
$(1-|s|^2)^{-1}$ approximately cancels for the lighter eigenvalue. The
cross-talk of the visible and the hidden sector gives rise to a
larger overlap of the light hidden state with the visible sector. Since
the $V_{2m}$ is orthogonal to $V_{1m}$ the overlap of the former state
with the visible sector becomes minimal in the limit where mixing in
the hidden sector dominates for $|s|\to 1$. In this limit the coupling
of $V_{1m}$ to SM matter saturates, while $V_{2m}$
decouples. \\

\begin{figure}[!t]
  \setcounter{figure}{7}
  \makeatletter
  \renewcommand{\thefigure}{\@arabic\c@figure a}
  \parbox{0.44\textwidth}
  {
    \vspace{-0.08cm}
    \includegraphics[height=5.5cm]{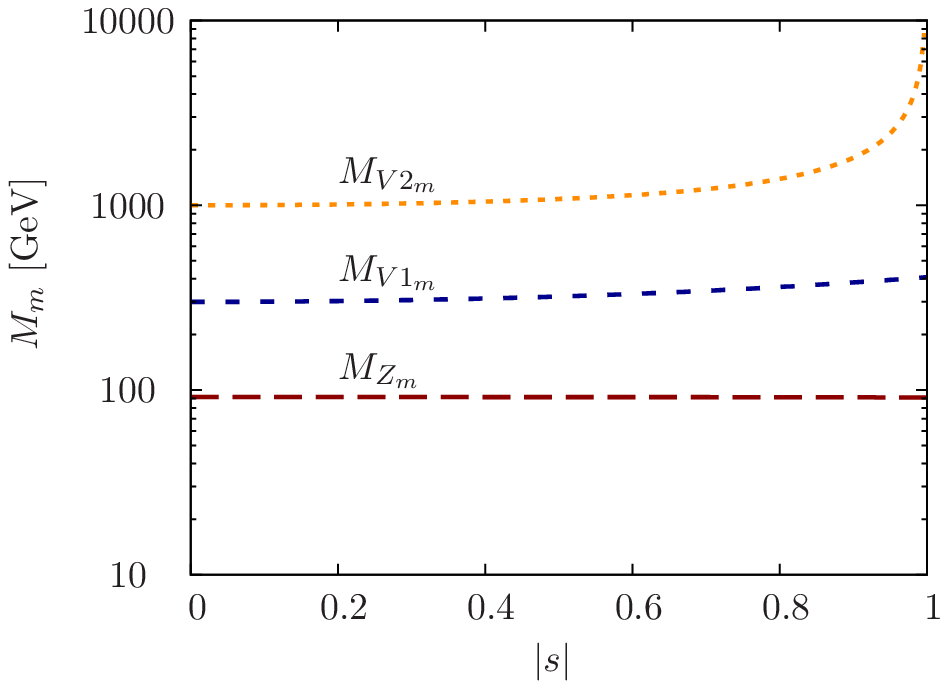}
    \vspace{0.08cm}
    \caption{\label{fig:12massv} Exact $V_m$ and $Z_m$ masses as a
      function of $|s|$ for
      $M_{V1_c},M_{V2_c}=300~{\text{GeV}},1000~{\text{GeV}}$.}
  }\hspace{0.8cm} \setcounter{figure}{7} \makeatletter
  \renewcommand{\thefigure}{\@arabic\c@figure b}
  \parbox{0.44\textwidth}
  {
    \includegraphics[height=5.5cm]{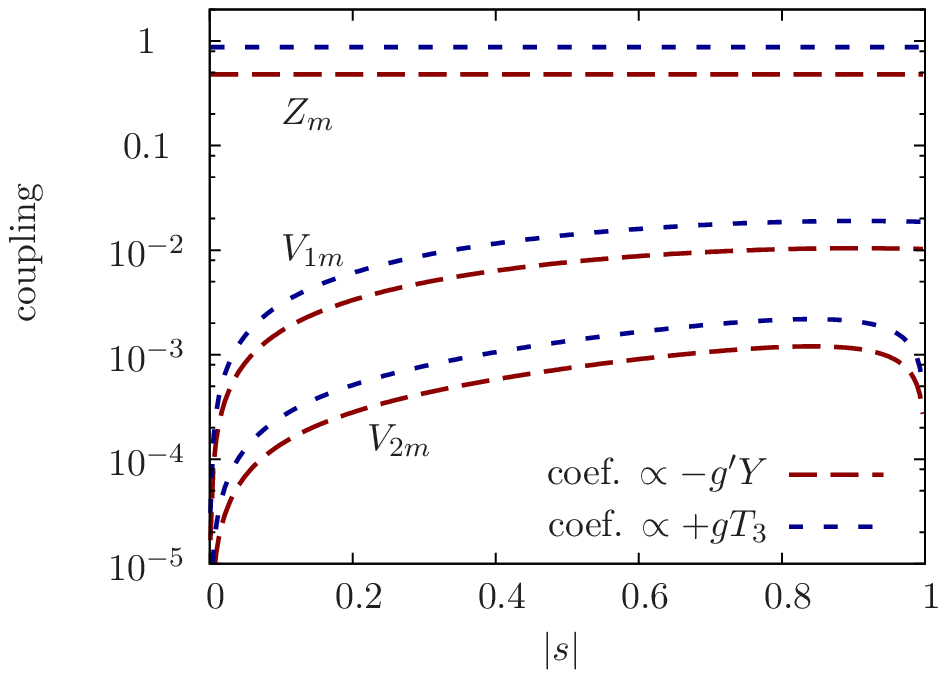}
    \caption{\label{fig:charge} Exact $V_m$ and $Z_m$ couplings as a
      function of $|s|$ for
      $M_{V1_c},M_{V2_c}=300~{\text{GeV}},1000~{\text{GeV}}$.}
  }
\end{figure}
\setcounter{figure}{8}
\makeatletter
\renewcommand{\thefigure}{\@arabic\c@figure}
%

The intertwining of the SM and hidden sectors may be illustrated by
noting the diagonal Higgs-vector couplings
\begin{eqnarray}
  H_{0m}\, W\,W / Z_m Z_m
  &=& \frac{2 M^2_{W_c / Z_c}}{v_0}\,\cos{\theta_{01}} \cos{\theta_{02}} \\
  H_{0m} V_{1m} V_{1m} / V_{2m} V_{2m}
  &=& \frac{2 M^2_{Z_c}s^2_W}{v_0}\,\sigma'^2|s|^2\,
  \cos{\theta_{01}} \cos{\theta_{02}} \, \cos^2\phi / \sin^2\phi
\end{eqnarray}
and the non-diagonal Higgs-vector couplings
\begin{eqnarray}
   H_{0m} Z_m V_{1m} /  Z_m V_{2m}
&=& \frac{2 M^2_{Z_c} s_W}{v_0}\, \sigma'|s|\,
    \cos{\theta_{01}} \cos{\theta_{02}} \, \cos\phi / \sin\phi \\
   H_{0m} V_{1m} V_{2m}
&=& \frac{M^2_{Z_c} s^2_W}{v_0}\, \sigma'^2|s|^2\,
    \cos{\theta_{01}} \cos{\theta_{02}} \, \sin 2\phi
\end{eqnarray}
in which the $W$ couplings are modified only by Higgs mixing while the
$Z_m$ and $V_{1m,2m}$ couplings are affected by the superposition of
Higgs and vector mixings. \\

\section{Summary}

\noindent
In this report we have taken a first modest step in analyzing
scenarios in which the SM is coupled to a hidden sector comprising more
than one degree of freedom. The difficulty of the analysis rises
enormously with the complexity of the hidden sector that can only be
accessed through mixing effects with the SM fields. The simplest
structures of the hidden sector are abelian or [extended] SM-type gauge
theories broken by the Higgs mechanism. The Higgs fields interact with
the SM Higgs field by means of bilinear quartic couplings, the vector
fields by means of kinetic mixing with the SM hypercharge field. \\

In the first part we have analyzed quite generally the mixing effects
within the Higgs system and within the vector system for an arbitrary
number of degrees of freedom. Closed and transparent analytical
solutions can only be obtained for $1 \oplus 1$
configurations. Approximate solutions however can be obtained quite
generally to leading non-trivial order of small mixings. In the Higgs
as well as gauge sector the mass matrix can
systematically be diagonalized to arbitrary order. \\

Two examples illustrate the abstract analysis. For the $1 \oplus 1$
system we have studied a complete set of observables which allow to
reconstruct the set of fundamental parameters in the Lagrangian, in
principle.  Higgs and vector-boson masses and their trilinear couplings,
supplemented by fermion couplings, generate such an ensemble. Since
all the Higgs and vector-boson states which are identified experimentally,
are mixed mass eigenstates, the effective vertices are complicated
mixtures in which Higgs mixing and kinetic mixing are intertwined. If
the mixing is too small the associated cross sections may, partly, not
be large enough to measure the vertices involved, but blocks of
essential elements
in the Lagrangian can nevertheless be isolated.\\

Precision analyses of the electroweak sector allow to constrain the
$1\oplus 1$ system at present and future colliders.  Direct detection
limits at the LHC in Drell-Yan type production will give upper
production limits on the additional neutral vector boson. Other
channels like $WW$ production and the search for anomalous $ZWW$
interactions will (and in fact already do) constrain the associated
mixing via modifying the SM coupling pattern, resulting in a
phenomenology sensitive to the described coupling modifications.
A precise measurement of the Higgs self-interaction  facilitates a
complete determination of the extended Higgs-sector
parameters. Current extrapolations suggest that only a lower limit on
the hidden-sector vacuum expectation value can be established.
The potentially clean environment of a lepton collider, however,
might be able to amend this conservative statement at high
luminosity.\\

The study has been extended to the system in which the SM interacts
with a hidden sector comprising two Higgs and vector-boson degrees
of freedom. It could be shown that for small mixing, in addition to
the numerical evaluation, this system also can be analyzed analytically.
Furthermore, the measurement strategies that we have outlined in
Sec.$\,$\ref{sec:1+1an} can be directly generalized to the $1 \oplus
2$ analysis. The situation, however, becomes less transparent due to
the increase in parameters while the number of phenomenologically
accessible measurements [especially in the Higgs sector where large
mixing effects could be present] stays the same.\\

{\it In toto.} If the Standard Model is coupled weakly to a complex
hidden sector, essential elements of this novel sector can be
reconstructed, though the experimental analysis may turn out very
difficult, and partly incomplete, if mixings are too small. \\

\setcounter{equation}{0}
\setcounter{section}{0}
\def\thesection{\Alph{section}}
\def\thesubsection{\Alph{subsection}}
\renewcommand{\theequation}{{\rm \thesubsection.\arabic{equation}}}

\section*{APPENDIX}

\subsection{Mixing in the Hidden Sector}
\label{append:exhiggs}
If quartic [bi-bilinear] mixing terms in the hidden sector are
included, the Higgs potential is generalized to
\begin{eqnarray}
 {\cal{V}}_{\cal{H}} &=& \sum_{i=0}^n \left[\, \mu^2_i |S_i|^2
                                    + \lambda_i |S_i|^4\, \right]
                        +\tfrac{1}{2} \sum_{i \neq j=0}^n \eta_{ij}\,
                                                    |S_i|^2 |S_j|^2
                                                    \nonumber \\
                    &=&  \left[\,\mu^{2T}\, |S|^2 + |S|^{2T}\, \lambda\, |S|^2\, \right]
                       +\tfrac{1}{2}\, |S|^{2T}\, \eta\, |S|^2            
\end{eqnarray}
in obvious vector/matrix notation in the second row.
The index $i = 0$ represents the Higgs field in the SM sector, i.e. $S_0 = \phi$ and
$\eta_{0j} = \eta_j\ \ [j=1,\ldots, n]$ etc, while indices $j \geq 1$ refer 
to hidden-sector scalar fields. \\

The visible and hidden components $v$ of the vacuum Higgs fields are
defined by the vanishing of the derivative of the Higgs potential,
\begin{eqnarray}
  v^2 &=& - \left\{ \lambda + \tfrac{1}{2} \eta \right\}^{-1} \mu^2  \nonumber \\
      &\simeq& - \left\{ \lambda^{-1} -  \tfrac{1}{2}
                                   \lambda^{-1} \eta \lambda^{-1}\, \right\}\, \mu^2
\end{eqnarray}
for small off-diagonal mixing parameters $\eta$ and ${\cal{O}}(1)$ diagonal parameters $\lambda$.  \\

The term bilinear in the physical fields defines the masses of the
Higgs particles,
\begin{eqnarray}
 {\cal{M}}_{H}^2 &=& 2\,v^T \left\{\lambda + \tfrac{1}{2} \eta \right\}\, v \nonumber \\[1.5mm]
                 &=& \left(\begin{array}{ccccc}
  2\lambda_0 v^2_0 &  \, \eta_1 v_0 v_1\,   & \, \eta_2 v_0 v_2\,   
                   &  \, \cdots\,           & \, \eta_n v_0 v_n\,  \\[1mm]
       {  }        & 2\lambda_1 v^2_1  &  \eta_{12} v_1 v_2   
                   & \cdots  & \eta_{1n} v_1 v_n  \\[2mm]
   \multicolumn{2}{c}{\{\,\rm symmetric\,\}}  &  \multicolumn{2}{c}{\ddots}  
                   &      \vdots \\[1mm]
       {  }        &      {  }         &   {     }         &  {   }  & 2\lambda_n v^2_n
        \end{array}       \right)  
\end{eqnarray}
Restricting the $n\times n$ symmetric mass matrix $M^2_c$ to the components of the hidden 
sector in the notation of Eq.$\,$(\ref{eq:m2}), the matrix can be 
diagonalized by an orthogonal transformation ${\cal{O}}_c$, modifying subsequently 
the phenomenological mixing vector $X$ in Eq.$\,$(\ref{eq:m3}):
\begin{eqnarray}
  M^2_c \, &\to& \,  M^2_{c/{\rm diag}} = {\cal O}_c M^2_c {\cal O}_c^T    \nonumber \\
  X     \, &\to& \, {\cal{O}}_c X     \,.
\end{eqnarray}
Notice that the off-diagonal ${\cal O}_c$ mixing elements change $X$ only  
to higher order so that the portal interactions between visible and hidden
sector fields are essentially not affected by the quartic mixing 
in the hidden sector. \\

Finally, the self-interactions among the Higgs fields, SM 
and hidden, can be derived from the potential 
\begin{equation}
  {\cal{V}}_{\cal{H}/{\rm self}} = \{ \lambda + \tfrac{1}{2} \eta \}_{ij}
                               \left[\, v_i H_{ic} H^2_{jc}
                             +   \tfrac{1}{4} H^2_{ic} H^2_{jc}\, \right]  
\end{equation}
in terms of the physical Higgs fields in the current [c]
representation.

\subsection{Dyadic Matrix}

It is straightforward to diagonalize the $n\times n$ dyadic matrix $D$
formed by a $n$-dimensional column vector $x=(x_1,\cdots,x_n)^T$ and its
transpose $x^T$ as
\begin{equation}
 D_{ij} = (x x^T)_{ij} =  x_i x_j \,.
\end{equation}
Making use of the rules for calculating determinants, one eigenvalue
emerges as positive and the other $(n-1)$ eigenvalues as zero:
\begin{eqnarray}
    d_1 = \sum^n_{i=1}\; x^2_i \ \ \mbox{and} \ \
    d_j = 0 \quad [\,j=2\;\, \mbox{to}\;\, n\,]    \,.
 \end{eqnarray}%
The eigenvectors associated with the eigenvalues read
\begin{eqnarray}
v_1 = d^{-1/2}_1\, x\,, \ \ \mbox{and}\ \
v_{j=2,..,n}
       \ \ \mbox{orthogonal}\;\; \mbox{to}\;\; v_1
\end{eqnarray}%
normalized to unity. \\

\setcounter{equation}{0}

\subsection{Block-diagonalization : Higgs and Vector Masses}

\noindent
{\bf 1.)} The eigen-masses and mixings in the {\it{\underline{Higgs sector}}},
when block-diagonalizing the real and
symmetric matrix ${\cal M}^2 \Rightarrow {\cal M}^2_m$ by an
orthogonal transformation ${\cal{O}}$,
\begin{eqnarray}
  \mathcal{M}^2
= \left( \begin{array}{cc}
         M_0^2     &  X^T     \\
             X     &   M^2
         \end{array}            \right)
=  \left( \begin{array}{cc}
                 0  & 0     \\
                 0  &   M^2
         \end{array}            \right)
+  \left( \begin{array}{cc}
              M_0^2 &  X^T     \\
              X     &  0
         \end{array}            \right) \quad \Rightarrow \quad
{\mathcal{M}}^2_m =  {\cal O} {\cal M}^2{\cal O}^T
= \left( \begin{array}{cc}
         {\hat{M}}_0^2    &   0   \\
           0              &   {\hat{M}}^2
         \end{array}            \right)
\end{eqnarray}
can iteratively be constructed from the lowest order to arbitrary
order in the expansion parameter $\epsilon \sim ||M^2_0|| / ||M^2||,
||X|| / ||M^2||$.  The first, large part of the mass matrix will be
called ${\mathcal{M}}^2_0$, the second, small part $\mathcal{E}$ which
is order $\epsilon$ compared with ${\mathcal{M}}^2_0$. Thus, the
expansion is valid for masses in the hidden
sector large compared to SM masses and the mixings. \\

The conditions which determine the mixing matrix $\mathcal{O}$ for
block-diagonalization of the mass matrix $\mathcal{M}^2$ are
orthogonality and diagonality:
\begin{eqnarray}
&&  \;\mathcal{O}
  = \sum_{N=0}^\infty  o_N \quad\; {\rm with} \quad\; o_0 = 1
  \quad\; {\rm and}  \quad\; o_N \sim \epsilon^N      \\[3mm]
&& {\mathcal{M}}^2_m  =
    \left( \begin{array}{cc}
            0      &   0   \\
            0      &   M^2
         \end{array}      \right) +
 \left( \begin{array}{cc}
                            M_0^2    &   0   \\
                              0      &   0
         \end{array}      \right) +
 \sum_{N=2}^\infty \left( \begin{array}{cc}
                        {\hat{M}}^2_{0,N}   &    0       \\
                         0                  &  {\hat{M}}^2_N
                   \end{array}          \right)
\end{eqnarray}
The first two matrices in ${\mathcal{M}}^2_m$ will occasionally be
identified
with indices $j = 0$ and 1, respectively. \\

\noindent {\it (i)} The {\it orthogonality} condition for
$\mathcal{O}$ determines the symmetric part of the component $o_N$
from $o_{j < N}$ as
\begin{eqnarray}
 o_N + o_N^T = - \sum^{N-1}_{j=1} o_j o_{N-j}^T    \,.
\end{eqnarray}

\noindent {\it (ii)} The {\it diagonalization} condition
of the mass matrix determines the antisymmetric part of $o_N$ from the
off-diagonal block elements, and at the same time the expansion of the
mass eigenvalues from the diagonal block elements:
\begin{equation}
\left(\begin{array}{cc}
                  {\hat{M}}^2_{0,N}  &  0             \\
                  0                  & {\hat{M}}^2_N
                  \end{array} \right)
         \,=\, \sum_{j=0}^{N} o_j\, \mathcal{M}^2_0\, o^T_{N-j}
          +\sum_{j=0}^{N-1} o_j\, \mathcal{E}\, o_{N-1-j}^T
          \ \ \mbox{with} \ \
{\cal M}^2_0 = \left(\begin{array}{cc}
                     0   &  0 \\
                     0   &  M^2
                     \end{array}\right)
          \ \ \mbox{and} \ \
{\cal E} = \left(\begin{array}{cc}
                     M^2_0   &  X^T \\
                     X       &  0
                     \end{array}\right)
\label{Eq:E}
\end{equation}
for $N\geq 1$.\\

To simplify the notation we switch from indexed symbols to one-letter
symbols by denoting
\begin{equation}
o_N  =  \left( \begin{array}{cc}
              x_N    &  y_N^T     \\
              -y_N   &  z_N
              \end{array}   \right)      \,.
\end{equation}
The matrix $z = z^T$ is taken symmetric, the antisymmetric part of
$o_N$ is defined in the $y$ column and row. To unify the mass
dimensions and express all the formulae in compact form, we introduce
three dimensionless and two dimensionful matrices as
\begin{eqnarray}
 \mu &=& M^{-2} M^2_0  \qquad\qquad\qquad\,
         y_{(N)+}  = M^2 y_{(N)}             \nonumber\\
 y   &=& M^{-2} X      \qquad\qquad\qquad\quad\!
         y_{(N)-}  = M^{-2} y_{(N)}          \nonumber\\
 \hat{z}_N &=& M^{-2} z_N M^2
\end{eqnarray}
where $y_\pm$ however always come as dimensionless combinations. \\

The simplified recurrence relations of the matrix blocks may be cast
in the following form for the block-diagonal components:
\begin{eqnarray}
\begin{array}{ll}
x_0 = 1
 & \hskip 0.5cm  z_0 = 1          \\
x_1 = 0
 & \hskip 0.5cm  z_1 = 0          \\
x_2  = - \tfrac{1}{2} y^T y
 & \hskip 0.5cm z_2
          =  -\tfrac{1}{2} yy^T  \\
x_3       =  - y^T  \mu y
  & \hskip 0.5cm
  z_3     = -\tfrac{1}{ 2} \{yy^T,\mu\}       \\
\hskip 0.6cm \vdots & \hskip 1.cm \vdots     \\
x_N = - \frac{1}{2} \Sigma^{N-1}_{j=1}
        \left(  x_j x_{N-j} + y_j^Ty_{N-j} \right)
  & \hskip 0.5cm z_N = -\tfrac{1}{2} \Sigma^{N-1}_{j=1}
        \left( z_jz_{N-j} + y_ jy_{N-j}^T \right) \ \      \\[1mm]
\end{array}
\end{eqnarray}
and for the off-diagonal components:
\begin{eqnarray}
&& \begin{array}{llll}
   y_0 = 0,         \ \   & \, \quad\,  y_1 = -y, \ \
 & \, \quad\, y_2 =  -\mu  y,  \ \
 & \, \quad\, y_3 = -\mu^2 y- \tfrac{1}{2} (y^T y y - y_+^T y y_-)
   \end{array} \nonumber\\
&& \hskip 0.7cm \vdots \nonumber\\
&& \hskip 0.1cm y_N =  \sum_{j=0}^{N-1}
   \left\{\, (\mu y_j - \hat{z}_j y)\, x_{N-1-j} - \hat{z}_{N-j}\, y_j
        + y_+^T\, y_{N-1-j}\, y_{j-} \, \right\}             \,.
\end{eqnarray}

The block-diagonal components of the mass matrix ${\cal M}^2_m$ are
given by
\begin{eqnarray}
&& \begin{array}{llll}
   {\hat{M}}^2_{0,0}     = 0            \,\qquad\,
 & {\hat{M}}^2_{0,1}     = M_0^2        \,\qquad\,
 & {\hat{M}}^2_{0,2}     = - y^T M^2 y  \,\qquad\,
 & {\hat{M}}^2_{0,3}     = -M^2_0 y^T y \ \ \ \ \\[1mm]
   {\hat{M}}^2_0 \;\,    = M^2          \,\qquad\,
 & {\hat{M}}^2_1 \;\,    = 0            \,\qquad\,
 & {\hat{M}}^2_2 \;\,    = \tfrac{1}{2}\{ yy^T, M^2\}  \,\qquad\,
 & {\hat{M}}^2_{3}\;\,   = M^2_0 y y^T \ \ \ \
  \end{array} \nonumber \\[1mm]
&& \hskip 1.0cm \vdots                                \nonumber\\[1mm]
&& \hskip 0.1cm {\hat{M}}^2_{0,N}
               = \sum_{j=0}^{N} y_j^T M^2 y_{N-j}
                +\sum_{j=0}^{N-1}\left[\,M^2_0\, x_j x_{N-1-j}
                              + (y^T_j X + X^T y_j\,) x_{N-1-j}\right]
                 \nonumber\\
&& \hskip 0.1cm {\hat{M}}^2_N \;\,
               = \sum_{j=0}^{N} z_j M^2 z_{N-j}\,
                +\,\sum_{j=0}^{N-1} \left[\, M^2_0\, y_j y^T_{N-1-j}
                             - (z_j X y^T_{N-1-j}
                               +y_{N-1-j} X^T z_j) \right]                \,.
\end{eqnarray}

\vskip 0.5cm
\noindent
{\bf 2.)} In the same way the mass matrix in the
{\it{\underline{gauge sector}}}
can be diagonalized recursively for small gauge-kinetic mixing $s$.
After applying the KT matrix $\cal{Z}$, given in closed form by
\begin{equation}
  {\cal{Z}}
= \left(\begin{array}{ccc}
        s_W    &   c_W         &  0        \\
        c_W    &  -s_W         &  0        \\
        0      &  -\sigma \,s   &  \sigma
         \end{array}\right)
\end{equation}
with the symmetric matrix $\sigma = (1-ss^T)^{-1/2}$, the transformed $(2+n)\times (2+n)$
mass matrix
\begin{equation}
    {\cal{M}}^2_s
  = {\cal{Z}}\, {\cal{M}}^2_c\, {\cal{Z}}^T
  = M^2_{Z_c}
    \left(\begin{array}{ccc}
           0     &     0              &   0                 \\[1mm]
           0     &     1              &  s_W  s^T \sigma   \\[1mm]
           0     &   s_W \sigma s     &   \sigma (\Delta + s^2_W s s^T)  \sigma
           \end{array}\right)
    \quad \mbox{with} \quad
    \Delta = M^2_{V_c}/M^2_{Z_c}
\end{equation}
has the characteristic properties which allow the recursive diagonalization
according to the algorithm developed in the previous subsection. Disregarding
the photonic null-vectors, we can identify, in symbolic notation,
\begin{equation}
  M^2_0 \sim M^2_{Z_c}\, ; \quad M^2 \sim M^2_{V_c} \,
                         ; \quad X \sim M^2_{Z_c} s
\end{equation}
for the $(1+n)\times (1+n)$ mass submatrix with
$|| M^2 || \gg M^2_0 \gg || X ||$. However, the kinetic mass matrix
includes additional $s$-dependent terms which can be expanded for
small $s$. They add contributions $\Sigma \, {\cal{E}}^k$ to the matrix
$\cal{E}$ in Eq.$\,$($\ref{Eq:E}$). Since they affect the matrix $o_N$
only by already known matrices $o_{j < N-1}$, they are easy to
incorporate. This procedure is straightforward, though technically
cumbersome, and we will not present the additional terms in detail. \\

\setcounter{equation}{0}
\subsection{Re-diagonalization}

\noindent
After the block-diagonalization the mass matrix ${\hat{M}}^2 = M^2 +
\Delta$ in the hidden sector is not diagonal anymore. Here, the correction
term $\Delta$ is of the second order or higher in $\epsilon$. It may
be re-diagonalized ${\hat{M}}^2 \rightarrow [{\rm {diag}}\, {\hat{M^2}}
]= M^2+\Delta^d$ by the orthogonal transformation $U = 1 + u$. Expanding
all the matrices, $\Delta^{(d)} = \sum \Delta^{(d)}_N$ and $u = \sum u_N$,
systematically in terms of the power $N\geq 2$ of $\epsilon$, the diagonal
mass matrix and the orthogonal transformation matrix can easily be
constructed recursively, as worked out before. \\

\noindent
The {\it orthogonality} condition determines the symmetric part of
$u_N$ in terms of the predetermined lower-order matrices $u_{\, k \leq N-2}$
by the relation
\begin{equation}
   u_N + u_N^T = - \sum^{N-2}_{k=2} u_k u^T_{N-k}
\end{equation}
Introducing the symmetric auxiliary matrix
\begin{equation}
   A_N = \sum_{k-2}^{N-2} \left[\, u_k M^2 u^T_{N-k}
                          +(u_{N-k}\, \Delta_k + \Delta_k\, u^T_{N-k})
       + \sum^{N-2-k}_{l=2} u_k\, \Delta_l\, u^T_{N-k-l}\, \right] \quad
\label{eq:diag}
\end{equation}
the {\it diagonalization} condition
\begin{equation}
   \Delta^d_N = \Delta_N + u_N M^2 + M^2 u^T_N + A_N
\end{equation}
can be exploited to project out the antisymmetric part of $u_N$,
\begin{equation}
   [u_N - u_N^T]_{ab} = 2 I_{ab} \left[\, \Delta_{N\,ab}
   + \tfrac{1}{2} (M^2_{aa}+M^2_{bb}) [u_N+u_N^T]_{ab}
    + A_{N\,ab} \right]
\quad [a \neq b]
\end{equation}
with the abbreviation $I_{ab} = 1 / (M^2_{aa} - M^2_{bb})$ and the
symmetric part calculated before by means of the orthogonality
condition.  The diagonalized eigenvalues are given by the elements of
the matrix $\Delta^d_N$ in Eq.$\,$(\ref{eq:diag}), which, at this
point, includes only predetermined matrices $u_N, A_N$ on the
right-hand side.
In this way the re-diagonalization of the mass matrix in the hidden sector is completed. \\

These solutions may be illustrated for the first three non-trivial
cases. The second- and third-order terms read
\begin{eqnarray}
N = {2,3}\, :  \qquad
 \Delta^d_{2,3\,aa}  &=&   \Delta_{2,3\,aa} \\
        u_{2,3\,aa}  &=&  0               \\
        u_{2,3\,ab}  &=& I_{ab}\, \Delta_{2,3\,ab} \quad [a\neq b]   \,.
\end{eqnarray}
The transformation matrix $u_{2,3}$ is apparently antisymmetric.
As a result, the newly diagonalized mass matrix in the hidden sector
is found by just truncating the mass matrix after block-diagonalization
to the diagonal elements up to the third order. However, starting
from the fourth order, there appear non-trivial contributions to
the diagonal elements from the lower-order terms. The fourth-order terms
read
\begin{eqnarray}
N = 4\, :  \qquad
 \Delta^d_{4\,aa} &=& \Delta_{4\,aa}
                   +\sum_{c\neq a} I_{ac}\, \Delta^2_{2\,ac} \\
 u_{4\,aa} &=& - \tfrac{1}{2}\,\sum_{c\neq a} I^2_{ac}\, \Delta^2_{2\,ac}\\
 u_{4\,ab} &=&  I_{ab}\, \Delta_{4\,ab}
              - I^2_{ab}\, \Delta_{2\,aa} \Delta_{2\,ab}
              + \sum_{c\neq a} I_{ab} I_{ac} \,
                               \Delta_{2\,ac}\, \Delta_{2\,bc}
                \quad [a\neq b]\,.
\end{eqnarray}
Note that the fourth-order matrix $u_4$ is not antisymmetric any more.

\vspace{8mm}
\noindent
{\bf Acknowledgements} \\

\noindent {\it C.E. thanks Joerg Jaeckel, Valya Khoze and Michael
  Spannowsky for helpful discussions.  The work of S.Y.C. was
  supported by Basic Science Research Program through the National
  Research Foundation (NRF) funded by the Ministry of Education,
  Science and Technology (NRF-2011-0010835).}




\end{document}